\documentclass{aa}

\usepackage{graphicx}
\usepackage{lipsum}
\usepackage[dvipsnames]{xcolor}
\usepackage{natbib}
\usepackage{hyperref}
\definecolor{myorange}{HTML}{ED9145}
\definecolor{myblue}{HTML}{21609D}
\hypersetup{%
colorlinks=true,
citecolor=myblue,
linkcolor=myorange,
 urlcolor=myorange,
linkbordercolor=myorange
}
\usepackage{algorithm2e}
\usepackage{array}
\usepackage{txfonts}
\usepackage{subfig}
\usepackage{threeparttablex}
\usepackage{longtable}
\usepackage{multirow}
\newcommand{\fink}{{\sc Fink}}
\newcommand{\elephant}{{\sc ELEPHANT}}

\newcommand\Tstrut{\rule{0pt}{2.6ex}}         % = `top' strut
\newcommand\Bstrut{\rule[-0.9ex]{0pt}{0pt}}   % = `bottom' strut

\usepackage{array,etoolbox}
\preto\tabular{\setcounter{magicrownumbers}{0}}
\newcounter{magicrownumbers}
\def\rownumber{}
\begin{document} 

%\title{Identifying Hostless Transients in High-Cadence Astronomical Surveys}
%\title{Less Hosts, More Hostless, even More Elephants}
\title{\includegraphics[scale=0.0175]{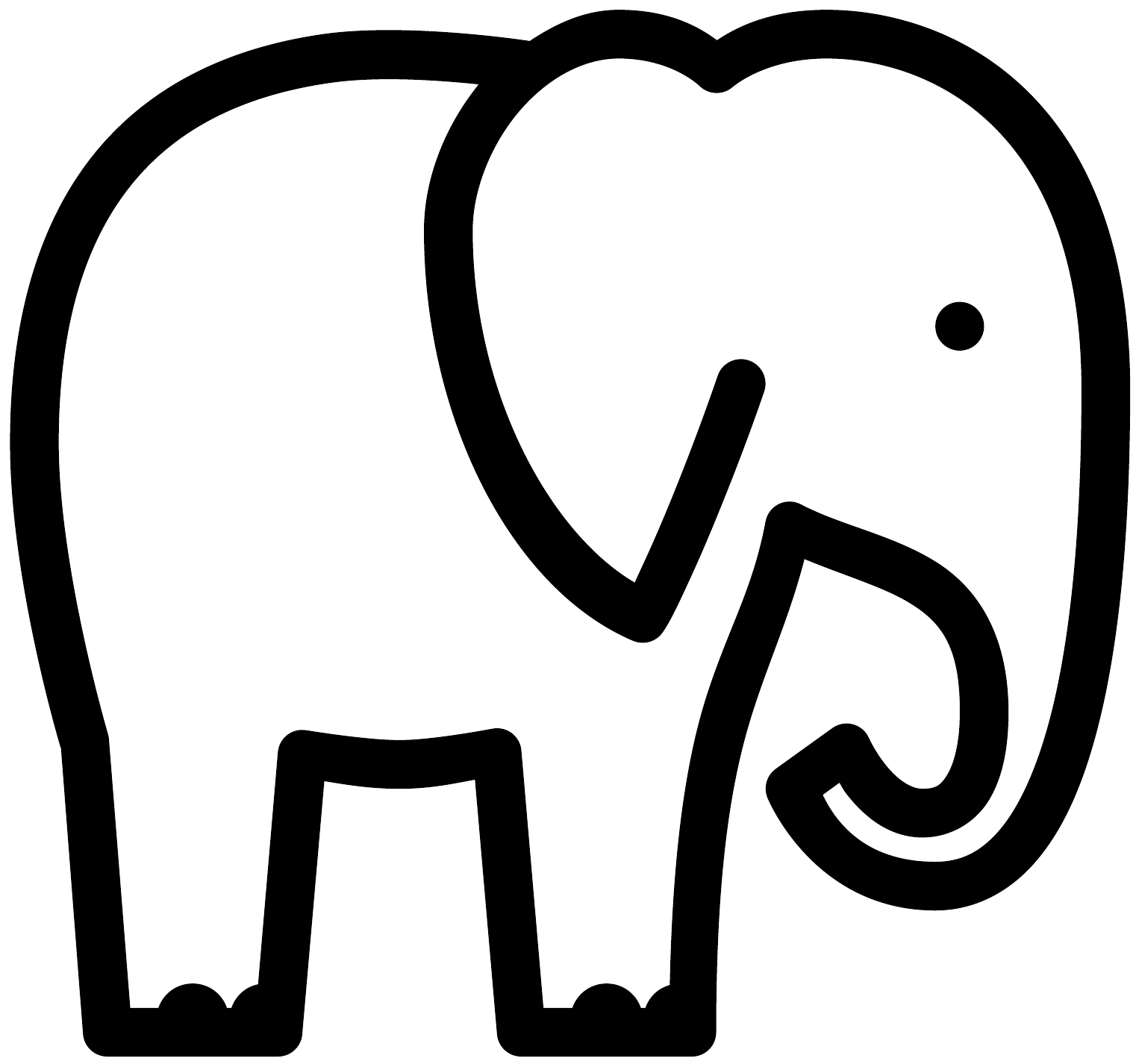}~\texttt{ELEPHANT}: ExtragaLactic alErt Pipeline for Hostless AstroNomical Transients
}

\author{P. J. Pessi\inst{1}\thanks{\email{priscila.pessi@astro.su.se}}
           \and
            R. Durgesh \inst{2} 
           \and
           L. Nakazono\inst{3} 
           \and
             E.~E. Hayes\inst{4}
            \and
            R.~A.~P. Oliveira \inst{5}
           \and 
            E.~E.~O. Ishida\inst{6}
            \and \\
            A. Moitinho\inst{7}
            \and
            A. Krone-Martins \inst{8} 
            \and
            B. Moews\inst{9, 10}
            \and 
           R.~S. de Souza \inst{11}
           \and
           R. Beck \inst{12}
            \and \\
           M.~A. Kuhn \inst{11}
           \and
            K. Nowak\inst{11}
            \and
            S. Vaughan\inst{13}
           (for the COIN collaboration)}

\authorrunning{Pessi et al.}

\institute{The Oskar Klein Centre, Department of Astronomy, Stockholm University, AlbaNova 106 91, Stockholm, Sweden 
\and 
Independent Researcher,  Ingolstadt, Germany 
\and 
Instituto de Astronomia, Geofísica e Ciências Atmosféricas da USP, 05508-900, São Paulo, Brazil
\and 
Institute of Astronomy and Kavli Institute for Cosmology, Madingley Road, Cambridge, CB3~0HA, UK 
\and
Astronomical Observatory, University of Warsaw, Al. Ujazdowskie 4, 00-478 Warszawa, Poland
\and
Universit\'e Clermont Auvergne, CNRS/IN2P3, LPC, F-63000 Clermont-Ferrand, France 
\and
CENTRA, Universidade de Lisboa, FCUL, Campo Grande, Edif. C8, 1749-016 Lisboa, Portugal
\and
Donald Bren School of Information and Computer Sciences, University of California, Irvine, CA 92697, USA
\and 
Business School, University of Edinburgh, 29 Buccleuch Pl, Edinburgh, EH8~9JS, UK
\and
Centre for Statistics, University of Edinburgh, Peter Guthrie Tait Rd, Edinburgh, EH9~3FD, UK
\and
Centre for Astrophysics Research, University of Hertfordshire, College Lane, Hatfield, AL10~9AB, UK
\and 
Independent Researcher,  Budapest, Hungary 
\and 
School of Mathematical and Physical Sciences, Macquarie University, NSW 2109, Australia
}

\abstract
% Context
{Transient astronomical events that exhibit no discernible association with a host galaxy are commonly referred to as hostless. These rare phenomena are associated with extremely energetic events, and they can offer unique insights into the properties and evolution of stars and galaxies. However, the sheer number of transients captured by contemporary high-cadence astronomical surveys renders the manual identification of all potential hostless transients impractical. Therefore, creating a systematic identification tool is crucial for studying these elusive events.}
% Aims
{We present the ExtragaLactic alErt Pipeline for Hostless AstroNomical Transients (\elephant),  a framework for filtering hostless transients in astronomical data streams. It was designed to process alerts from the Zwicky Transient Facility (ZTF) as presented in the Fink broker; however, its underlying concept can be applied to other data sources. 
}
%Methods
{We used Fink to access all the ZTF alerts produced between January/2022 and December/2023, selecting alerts associated with extragalactic transients reported in SIMBAD or TNS, as well as those classified as supernova (SN) or kilonova (KN) by the machine learning (ML) classifiers within the broker.  We then processed the associated stamps using a sequence of image analysis techniques to retrieve hostless candidates.}
% Results
{We find that $\lesssim$ 2\% of all analyzed transients are potentially hostless. Among them, only $\sim$ 10\% have a spectroscopic class reported on TNS, with Type Ia supernova being the most common class, followed by superluminous supernova. In particular, among the hostless candidates retrieved by our pipeline,  there was SN~2018ibb, which has been proposed to be a Pair Instability SN candidate; and SN~2022ann, one of only five known SNe~Icn.  
When no class is reported on TNS, the dominant classes are QSO and SN candidates, with the former obtained from SIMBAD and the latter inferred using the Fink ML classifier.}
% Conclusions
{\elephant\ represents an effective strategy to filter extragalactic events within large and complex astronomical alert streams. There are many applications for which this pipeline will be useful, ranging from transient selection for follow-up to studies of transient environments.  The results presented here demonstrate the feasibility of developing specially crafted pipelines that enable a variety of scientific studies based on large-scale surveys. \elephant\ is publicly available in the COINToolbox: \url{https://github.com/COINtoolbox/extragalactic_hostless}.}
\keywords{Methods: data analysis -- Astronomical databases: miscellaneous -- Stars: general -- Methods: statistical}

\maketitle

%%%%%%%%%%%%%%%%%%%%%%%%%%%%%%%%%%%%%%%%%%%%%%%%%%

%%%%%%%%%%%%%%%%% BODY OF PAPER %%%%%%%%%%%%%%%%%%

\section{Introduction}

Contemporary wide-field, untargeted surveys that scan large portions of the sky on a regular basis, such as the All-Sky Automated Survey for Supernovae \citep[ASAS-SN,][]{2014AAS...22323603S}, Gaia \citep{2016A&A...595A...1G}, the Asteroid Terrestrial-impact Last Alert System \citep[ATLAS,][]{2018PASP..130f4505T} and the Zwicky Transient Facility (ZTF; \citealt{2019PASP..131a8002B}), have significantly increased the number of transients discovered nightly over the past decades\footnote{See \cite{2017hsn..book...29Y} for numbers on the growth of discovered and classified supernovae from 1991 to 2015. For statistics on transient discovery and classification from 2016 onward, refer to \url{https://www.wis-tns.org/stats-maps}.}. Such projects have not only increased the number of confirmed transients of known classes but have also facilitated the discovery of new classes of events \citep[e.g.,][]{2014ApJ...794...23D, 2017NatAs...1..865K}. Thus, the past decade has witnessed a significant increase and diversification of the transient sky landscape, populated by a myriad of objects \citep[e.g.][]{2023PASP..135j5002H}.

Extragalactic transients can be described as the observational consequence of energetic events taking place outside the Milky Way. This description implies a progenitor population of astrophysical sources which should, in principle, be associated to a host galaxy. Nevertheless, a small fraction of transients seem to not be associated to any host and are thus considered \textit{hostless} \citep[e.g.][]{2022ApJS..259...13Q,2024MNRAS.tmp..936Q}. 
In these cases, the host may remain undetected either because it is fainter than the survey's limiting magnitude or because the transient was produced by a progenitor that achieved hypervelocity and escaped its host galaxy \citep[e.g.,][]{2006AJ....131.3047M,2011A&A...536A.103Z}. 
Hostless transients have been associated with superluminous supernovae \citep[SLSNe; e.g.][]{2015MNRAS.448.1206M}, gamma-ray burst \citep[GRBs; e.g.][]{2020ApJ...905...98H}, Fast X-ray transients \citep[FXTs; e.g.][]{2024arXiv240410660G} and lensed transients \citep[e.g.][]{2020MNRAS.495.1666R}, among others. 
Independently of the exact mechanism that rendered them hostless, such rare events represent an opportunity to further investigate peculiar astrophysical scenarios and may provide important clues regarding their local environment. 
They have already been used to discover low surface brightness galaxies \citep[LSB,][]{2012A&A...538A..30Z} and to study intra-cluster stellar populations \citep[][]{2015ApJ...807...83G}.  

Given such scientific potential, whenever a hostless transient is discovered, it sparks the interest of the astronomical community focused on rare events. In the past, the moderate number of discovered transients allowed thorough investigation of each candidate together with their associated hosts \citep[e.g.,][]{1997ARA&A..35..309F,1981PASP...93....5B}. 
Nowadays, untargeted searches are discovering transients in fainter and more distant host galaxies, substantially increasing their numbers and rendering it impossible to study all of them in detail. As an example, ZTF currently detects a few hundred thousand transient candidates per night, while the upcoming Vera C. Rubin Observatory Legacy Survey of Space and Time (LSST) is expected to detect around 10 million per night over a period of 10 years \citep{ldm612}. In this context, it became necessary to develop automated frameworks for mining large astronomical datasets.

In this work, we introduce the ExtragaLactic alErt Pipeline for Hostless Transients (\elephant), whose goal is to enable automatic identification of confirmed or potential extragalactic events without an obvious host association. It significantly reduces the number of candidates requiring visual inspection, thus allowing an optimal allocation of expert time and follow-up resources. 
\elephant\ employs a range of established image processing techniques to analyze image stamps associated with each transient, assessing the likelihood of a host's presence. We detail the components of our pipeline and discuss a number of noteworthy candidates identified during its development. We visually inspected candidates with an associated spectroscopic classification available on the Transient Name Server (TNS\footnote{\url{https://www.wis-tns.org/}}) to confirm their hostless nature. This process also helped us define statistical thresholds to apply to the rest of the sample. We found that less than $\lesssim$ 2\% of the analyzed sample is potentially hostless, with the most common classes of hostless candidates being QSOs, Type Ia SN, and SLSN. Some hostless candidates identified by our pipeline, which present interesting features, had already been thoroughly discussed in the literature  (see Section~\ref{sec:res}). Our results illustrate the potential of the pipeline if applied to more recent data. We are currently working in integrating it to the Fink broker \citep{Moller2021}, which will allow processing ZTF alert stream in real time and increase the chances of identifying hostless transients while they are still bright enough for spectroscopic follow-up.

This paper is organized as follows: Section~\ref{sec:data} outlines the data selected for this analysis. Section~\ref{sec:pipe} describes the \elephant\ workflow. Results are presented in Section~\ref{sec:res} and conclusions in Section~\ref{sec:conc}.

\section{Data}
\label{sec:data}

We use image data available within alerts distributed by ZTF. An alert package is produced when the difference imaging pipeline identifies a transient source. It includes photometric history, metadata, and three stamps: the original reference image -- \textit{template}, the new observation -- \textit{science},  and the difference image -- \textit{difference}  \citep{2019PASP..131a8002B,2019PASP..131a8003M}. This information is distributed nightly to community brokers, whose task is to filter, add value, and redistribute the alerts to domain experts. This work uses the alert stream information as provided by the Fink broker \citep{Moller2021}, however the pipeline is flexible enough to be used with other data sources\footnote{Other known community brokers include \texttt{ALERCE} \citep{2021AJ....161..242F}, \texttt{AMPEL} \citep{ampel}, \texttt{ANTARES} \citep{antares}, \texttt{Babamul}, \texttt{LASAIR} \citep{lasair} and \texttt{Pitt-Google}.}.

We retrieved all alerts processed by Fink between January/2022 and December/2023. The data set contained 70 176 557 alerts, which correspond to 17 683 691 objects. Approximately 50\% of these have an associated classification. We only keep events associated with an extragalactic transient classification, including all classes of active galactic nuclei (AGN), supernovae (SNe), and kilonova candidates, among others (the complete list of the classes considered for this work can be found in our repository\footnote{\url{https://github.com/COINtoolbox/extragalactic_hostless}}). The classifications provided by Fink are obtained via cross-match with  SIMBAD\footnote{\url{https://simbad.cds.unistra.fr/simbad/}} \citep{simbad}, the Transient Name Server\footnote{\url{https://www.wis-tns.org/}} (TNS), or produced by machine learning (ML) algorithms used by the broker \citep{moller2020, leoni2022, biswas2023}. In case a cataloged classification is available, we consider it to be final. However, ML-based classifications are given per alert. Since one object can produce many alerts, this sometimes results in different classes associated with the same astrophysical source. When selecting sources for which only ML classification is available, the final class was chosen by majority vote, taking into account all alerts associated with the same object. 

We exclude alerts with no associated classification or associated with galactic transients such as variable stars or objects present in the Minor Planet Center\footnote{\url{https://minorplanetcenter.net/about}}. Since we are only interested in hostless events, we also considered cross-match with the MANGROVE catalog \citep{ducoin2020} and removed any object associated with a known host, even if the host galaxy association is tentative. 
We keep only $\sim$ 3.5\% of the original alerts by applying these conditions. 
To eliminate potentially bogus events, we only consider transients with two or more alerts, meaning that they will have more than one associated set of stamps. 
The stamps are typically $63 \times 63$ pixels with the detected transient located at the center. Smaller alerts are produced in rare cases, normally related to detector edge effects or due to defects in the image acquisition process \citep{reyes2023}. 
To guarantee a homogeneous sample, we removed any stamp whose size is smaller than the typical value.
After applying these last conditions, we end with a total of 90 928 transients.

\begin{figure}
	\includegraphics[width=\columnwidth]{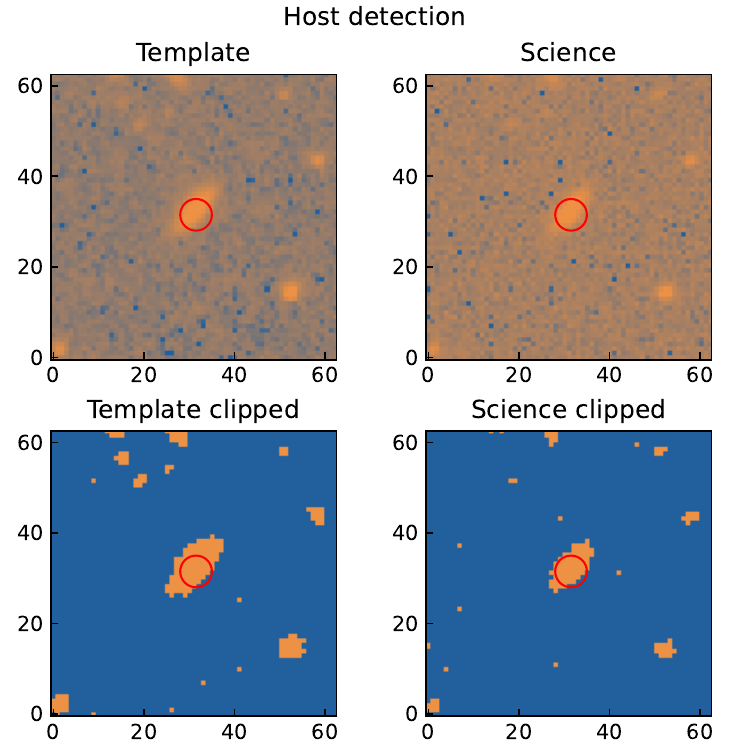}
    \caption{Example of the \textit{template} (left) and \textit{science} (right) stamps for a transient associated with a host galaxy. The top row shows the original stamps and the bottom row shows the masks produced from sigma clipping. At the center of the stamps, we display a red circle of 7 pix radius that indicates the aperture radius of the associated photometry.} 
    \label{fig:succeshost}
\end{figure}

\section{The \elephant\ Pipeline}
\label{sec:pipe}

The pipeline analyses both the \textit{science} and \textit{template} stamps in parallel. Thus, a source is considered hostless if either its \textit{template} or \textit{science} stamps survives all filtering stages. In principle, the \textit{template} image should suffice to detect the presence of a possible host, however, because of the \textit{template} generation process \citep[see][]{2019PASP..131a8003M}, some of them can suffer from transient contamination. In these cases, the transient would be detected as a source in the center of the \textit{template} image, leading to the wrong detection of a host. Considering both the \textit{template} and \textit{science} stamps attenuates this issue. 
Below, we describe each step of the pipeline. 

\subsection{Stamp pre-processing}
\label{sec:prepros}

If a stamp contains pixels with missing or empty values, the pipeline estimates the probability density function (PDF) of the counts in the remaining pixels via Gaussian resampling using the \texttt{scipy.stats.gaussian\_kde} Python method. The empty value is then replaced by randomly selected values from the resulting PDF, producing a homogenized sample where all images have the same number of valid pixels.
Additionally, we use the full-width at half maximum (FWHM) of each stamp to estimate the image quality. 
In our sample, the FWHM can vary from $\rm{FWHM} < 1.0^{\prime\prime}$ (a few cases) to $\rm{FWHM} > 3.0^{\prime\prime}$, with a median value of $\rm{FWHM} \sim 2.0^{\prime\prime}$. 
To select only the best available images representing each astrophysical source, all alerts associated with a given source are separated into 3 FWHM bins: $\rm{FWHM} < 1.0^{\prime\prime}$, $1.0^{\prime\prime}<\rm{FWHM}<2.0^{\prime\prime}$ and $\rm{FWHM} > 2.0^{\prime\prime}$. The pipeline only considers the stamps in the smallest available FWHM bin for each source, discarding all others.

All selected stamps for a given object are then stacked by adopting the median count value in each pixel of the $63 \times 63$ cutout.
This stacking process aids to enhance the images' signal-to-noise ratio (S/N), thereby improving the identification of potential hosts. Since the \textit{science} stamps result from a single exposure, this process impacts them much more than their \textit{template} counterparts. 
Nevertheless, this technique also serves to homogenize the effects of varying \textit{templates} used throughout the lifespan of a given transient.
 
\subsection{Segmentation masks}
\label{sec:segmentation}

\begin{figure}
	\includegraphics[width=\columnwidth]{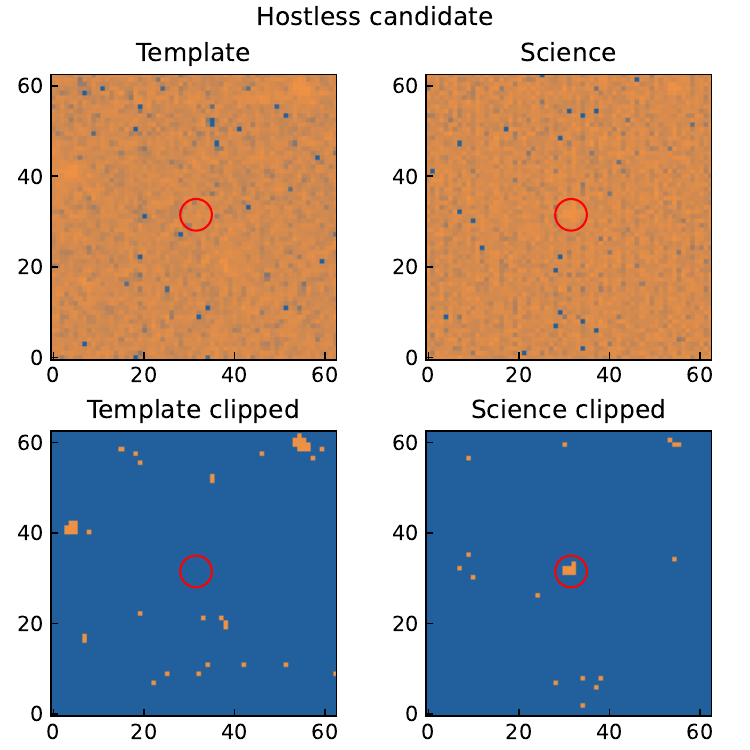}
    \caption{Example of the \textit{template} (left) and \textit{science} (right) stamps of a hostless transient candidate. The top row shows the original stamps, and the bottom row shows the masks produced by sigma clipping. We can see that the \textit{science} stamp shows a mask at the center of the stamp that is absent in the \textit{template} stamp. The absence of a mask is considered as the absence of a host. At the center of the stamps, we display a red circle of 7 pix radius that indicates the aperture radius of the associated photometry.}
    \label{fig:succesnohost}
\end{figure}

\elephant\ uses sigma clipping to mask sources present in the stamps and uses those masks to detect the presence of a host galaxy. 
Sigma clipping is a typical method to detect outliers in astronomical images, usually used to remove the effect of defective pixels or cosmic rays by clipping out pixels above a given sigma threshold. The values of the clipped pixels can then be replaced with a mask or filled in with some characterization of the remaining image counts. 

The ZTF alert package includes the aperture magnitude of the transient obtained from aperture photometry, calculated considering a 7-pixel radius aperture. We use this size as a reference for the maximum size of any detected transient. 
\elephant\ implements the \texttt{astropy.stats.sigma\_clip}\footnote{\url{https://docs.astropy.org/en/stable/api/astropy.stats.sigma_clip.html}} Python method considering $\sigma = 3$, median as the statistic to compute the clipping center value, and a maximum of ten iterations. 
As a result, any pixels above the selected median threshold are clipped. The clipped segments of the stamp are considered as the mask. If a mask bigger than 5 continuous pixels is found at the center of the \textit{science} stamp but not at the center of the corresponding \textit{template} stamp, or vice-versa, we flag the transient as a potential hostless candidate. 

\elephant\ utilizes the obtained masks to identify the position of the pixel closest to the center that corresponds to a detected neighboring mask, considering any masked pixel within a 7-pixel square as indicative of a neighbor's presence. Details on how the distance is computed can be found in Appendix~\ref{app:neighbor}. Although we don't further use the distance information here, a future user could consider it to additionally assess the presence of a host. This could be useful when analyzing SN, as they could occur on the outskirts of their hosts. In such a case, a mask will not be found at the center of the stamp but close to it. In this context, what is considered to be close should be defined by the user. Another popular image segmentation software in astronomy is \texttt{SExtractor}; we decided not to use it here as it requires more resources than sigma clipping, and it also requires the pipeline to use out-of-memory processing; for further discussion on the use of \texttt{SExtractor} see Appendix~\ref{app:sex}.

\begin{figure}
	\includegraphics[width=\columnwidth]{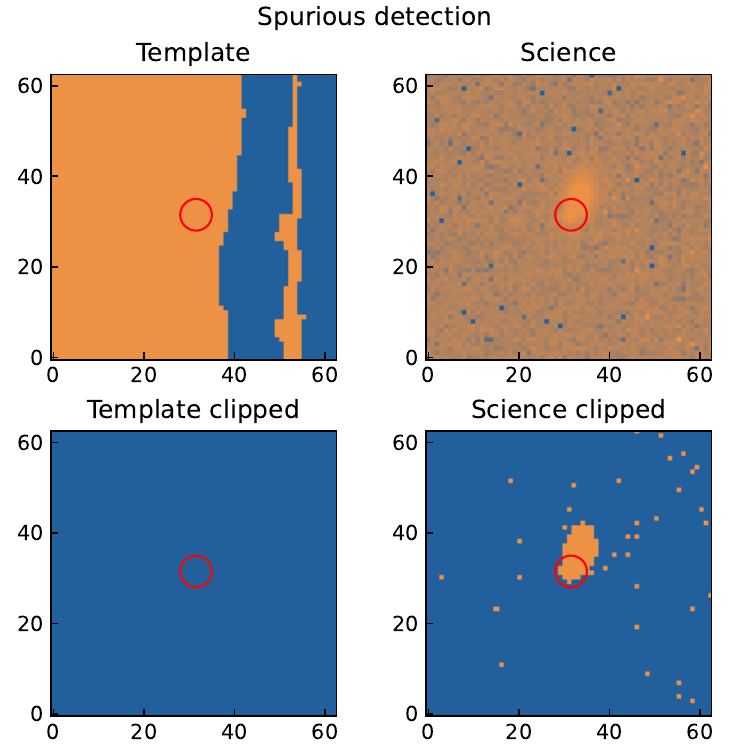}
    \caption{Example of the \textit{template} (left) and \textit{science} (right) stamps of a spurious hostless candidate detection. The top row shows the original stamps and the bottom row shows the masks produced from sigma clipping. We can see that the erroneous detection is driven by artifacts present in the original \textit{template} stamp. As a result, sigma clipped \textit{template} shown on the bottom left panel shows no signal. At the center of the stamps, we display a red circle of 7 pix radius that indicates the aperture radius of the associated photometry.}
    \label{fig:fail}
\end{figure}

After applying sigma clipping, \elephant\ retrieves 1669 hostless candidates. Fig.~\ref{fig:succeshost} and Fig.~\ref{fig:succesnohost} show an example of a host detection and of the detection of a hostless candidate, respectively. Fig.~\ref{fig:succeshost} shows that the presence of a host galaxy at the center of the stamp is seen as a mask in the center of both the \textit{template} and \textit{science} stamps. On the other hand, Fig.~\ref{fig:succesnohost} shows that a mask is present at the center of only one of the transient's stamps thus, it is flagged as a hostless candidate. Fig.~\ref{fig:fail} shows a spurious detection of a hostless candidate. In this case, the erroneous detection is driven by artifacts present on the \textit{template} stamp.

\subsection{Host categorization via Fourier power spectrum}
\label{sec:Fourier}

To further examine the presence or absence of a host, if a transient is flagged as a hostless candidate by the sigma clipping method, \elephant\ explores the Fourier space projections of the masked stamps. 
This strategy is reflective of methodologies previously applied to the classification of natural images across various landscapes \citep{PS2003}. By transforming the stamps into Fourier space, the pipeline is able to search for correlations in the background noise that can suggest the presence of a faint host, which would otherwise not be detected via the sigma clipping approach. 
This process involves calculating the medianized 1-dimensional  power spectrum from the 2-dimensional Fourier transform of the images. 
The mathematical foundation of this method is laid out as follows: the Fourier transform, denoted by $F(u, v)$, of an image, $I(x, y)$,  is calculated according to:
\begin{equation}
F(u, v) = \mathcal{F}\{I(x, y)\},
\end{equation}
where $(x, y)$ represents the pixel coordinates and $(u, v)$ the frequency domain coordinates. From this, the power spectrum, $P(u, v)$, is derived through the equation:
\begin{equation}
P(u, v) = |F(u, v)|^2.
\end{equation}
The median power, $M(k)$, for each radial frequency $k = \sqrt{u^2 + v^2}$, is calculated by taking the median of the power values across all angular coordinates $\theta$ for a given power $k$:
\begin{equation}
M(k) = \mathrm{median}\{P_k\}.
\end{equation}
We assume that the power spectrum of an image containing even a faint host signal will distinguish itself from the power spectrum of another from which sources were removed and whose pixels have been randomly shuffled, and consequently does not contain any spatially coherent information to be extracted.

\begin{figure*}
    \begin{minipage}{0.7\textwidth}
    \includegraphics[width=\textwidth]{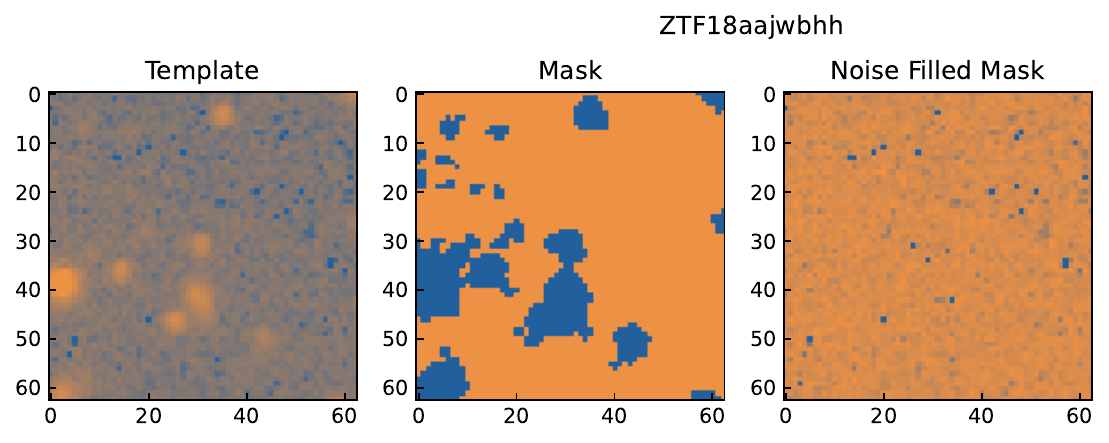}
    \end{minipage}
    \begin{minipage}{0.3\textwidth}
    \vspace{1cm}
    \includegraphics[width=0.85\textwidth]{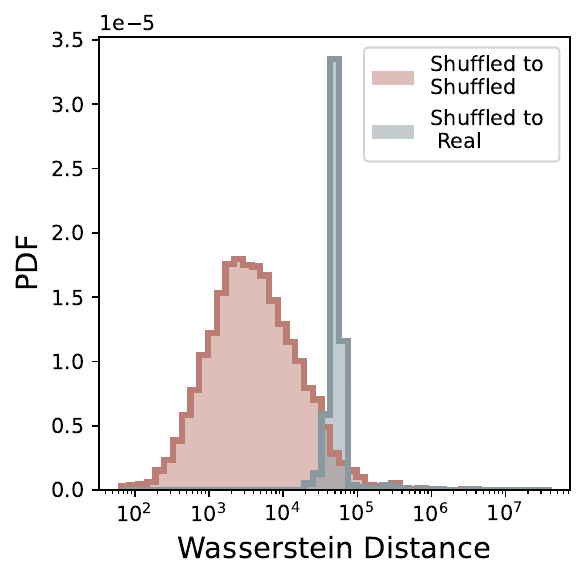}
    \end{minipage}
    \caption{Stages of the power spectrum analysis for a template with host (SN2017iuu / \texttt{ZTF18aajwbhh}). From left to right the panels show the template image, the mask and the mask populated with noise. The right-most panel shows the distribution of Wasserstein distances between the original template and shuffled noised masks (gray) and between random pairs of shuffled noised masks (rose). The distributions were generated using 1000 different shuffles of the noised masks within the central patch of 7 $\times$ 7 pixels. }
    \label{fig:stamps_whost}
\end{figure*}

\begin{figure*}
    \begin{minipage}{0.7\textwidth}
        \includegraphics[width=\textwidth]{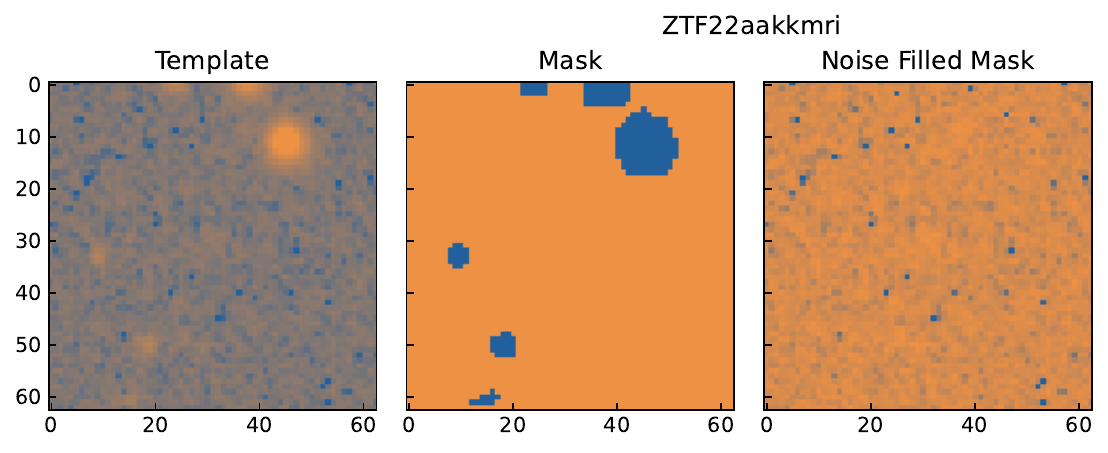}
    \end{minipage}
    \vspace{-0.05cm}
    \begin{minipage}{0.3\textwidth}
        \vspace{1cm}
        \includegraphics[scale=0.505]{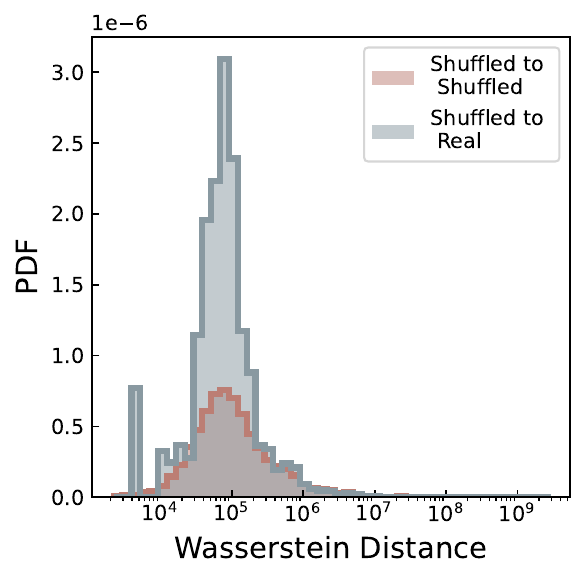}
    \end{minipage}
    \caption{Stages of the power spectrum analysis for a hostless template (SN2022knm / \texttt{ZTF22aakkmri}). Panel descriptions are equivalent to those described in Figure \ref{fig:stamps_whost}.}
    \label{fig:stamps_nohost}
\end{figure*}

To explore this, we first compute the power spectrum of the original image. Subsequently, we use the masks resulted from sigma clipping and fill masked sections with random noise sampled from the pixel value distribution of the masked image itself. The images are then cropped to three distinct sizes: 7 $\times$ 7, 15 $\times$ 15, and 29 $\times$ 29 pixels, always with the center coinciding with the position of the transient. Afterwards, we randomly shuffle the pixel positions and the power spectrum is recalculated. This process is repeated 1000 times. The radially averaged 1D power spectrum of the original image is then compared to those of each shuffled iteration using the Wasserstein distance, $W(p, q)$:

\begin{equation}
W(p, q) = \inf_{\gamma \in \Pi(p, q)} \int_{X \times Y} \|x - y\| \, d\gamma(x, y), 
\end{equation}

\noindent which measures the distance between the $p$ and $q$ distributions. 
The presence of a host, even if weak, is suggested if the distances from the original image's power spectrum to those of the shuffled images are on average greater than the distances between the power spectra from shuffled images themselves (see right panel of Fig~\ref{fig:stamps_whost} for an example of distance distributions when a host is present, and Fig~\ref{fig:stamps_nohost} for an example of the distance 
 distributions for a hostless candidate). 

This process yields a sample of 1000 distances for comparisons between the original image's power spectrum and the power spectra of the shuffled images for each cutout size. The final step involves estimating the Kolmogorov-Smirnov (K-S) statistic to quantify the similarity between these two distributions of distances. The K-S statistic is calculated using the following equation:
\begin{equation}
D = \sup_{x \in \mathbb{R}} | S_{1}(x) - S_{2}(x) |,
\end{equation}
where $D$ quantifies the maximum discrepancy between the cumulative distribution functions (CDFs) of two distinct samples. Here, $S_{1}(x)$ represents the empirical cumulative distribution function (ECDF) for the first sample, which consists of the Wasserstein distances between the power spectrum of the original image and those derived from shuffled images. $S_{2}(x)$, on the other hand, corresponds to the ECDF of the second sample, namely the distribution of distances among the shuffled images themselves. We use $D$ as a proxy for identifying the presence of a faint host in all images which survived the sigma clipping selection.  

\section{Results}
\label{sec:res}

\elephant\ combines 2 stages of filtering. All objects flagged as potential hostless candidates by the sigma clipping step (Section \ref{sec:segmentation}) were submitted to the power spectrum analysis (Section \ref{sec:Fourier}). This last stage attached to each object a K-S statistic value, $D$, which was constructed as a proxy indicating the presence of a faint host. We used a subset of visually inspected objects to define a selection cut threshold based on $D$ (Section \ref{subsec:threshold}), and analyzed the results from imposing such a threshold on a subset of spectroscopically confirmed transients (Section \ref{subsec:objs}).

\subsection{$D$ threshold for hostless candidates}
\label{subsec:threshold}

\begin{figure}
	\includegraphics[width=\columnwidth]{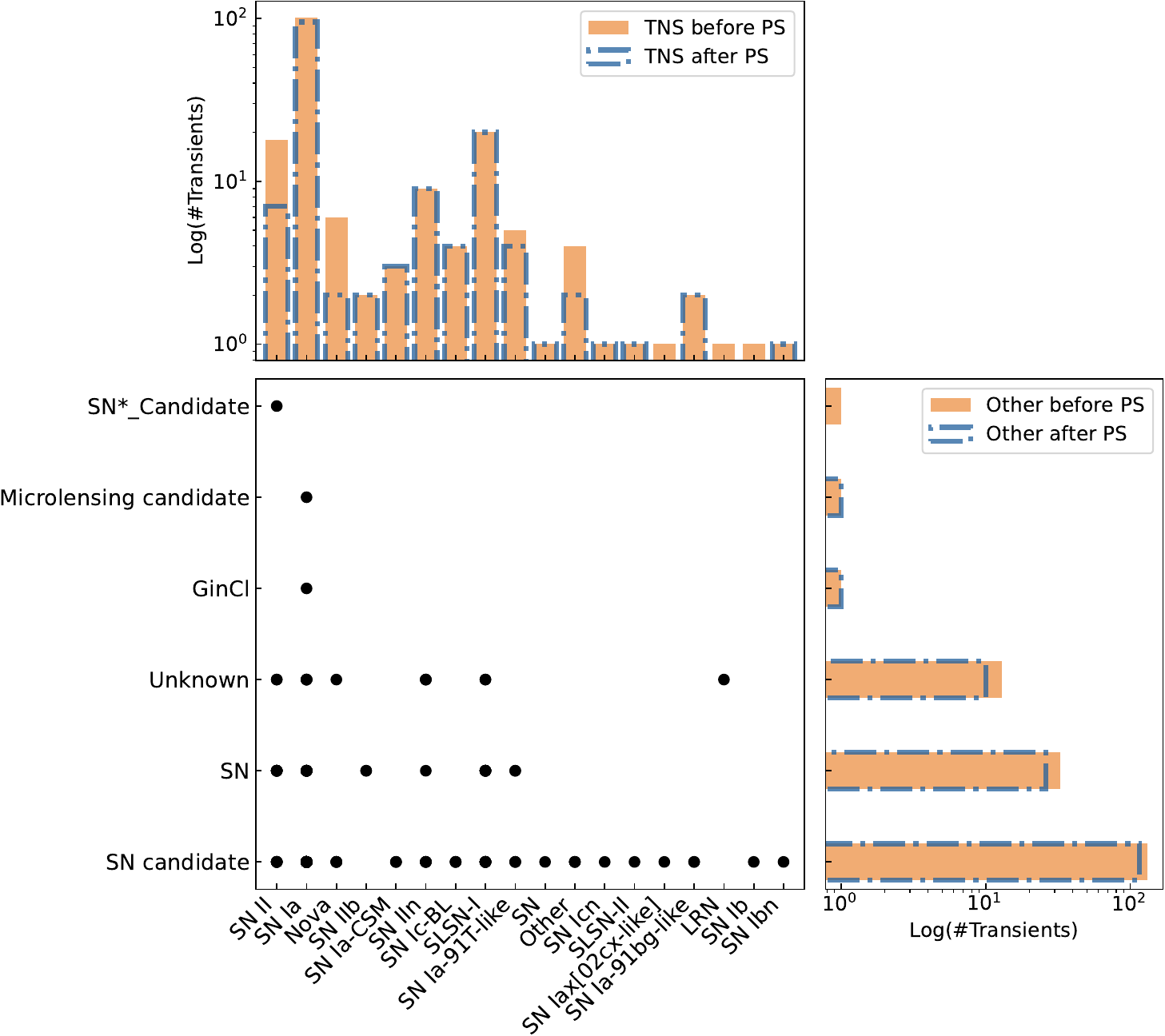}
    \caption{Central panel: comparison between spectral classification reported on TNS (horizontal axis) and the classification reported by Fink obtained from other sources (vertical axis). The $x$ and $y$ axis side panels show the number of transients considered to be hostless by the sigma clipping method before applying the power spectrum (PS) analysis (orange), and the number of surviving hostless candidates after applying the PS analysis (blue).}
    \label{fig:TNSclass}
\end{figure}

\begin{figure*}
    \centering
    \includegraphics[width=\textwidth]{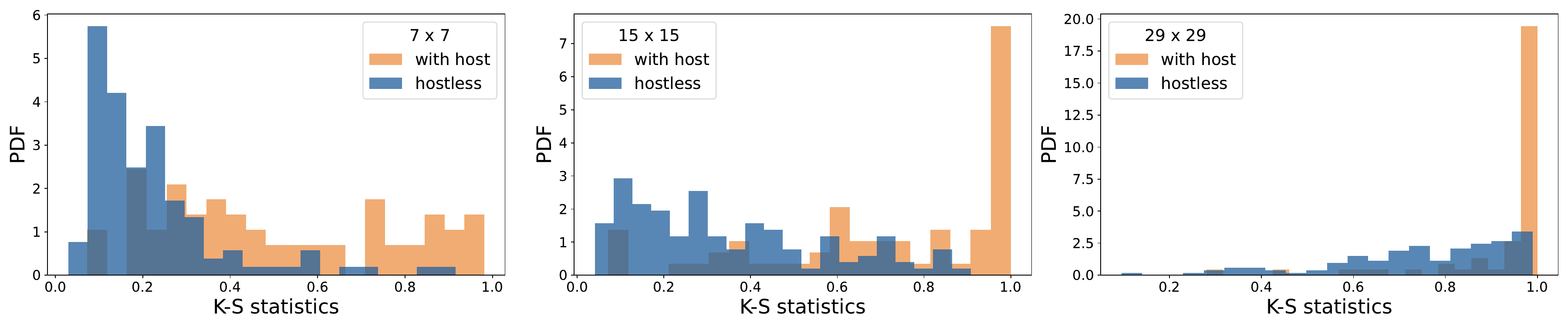}
    \caption{Distributions of the Kolmogorov-Smirnov statistic for the 181 objects with TNS classifications. The two categories, with host (orange) and hostless (blue) were identified through visual inspection. Panels show distributions obtained through the power spectrum analysis (Section \ref{sec:Fourier}) for different image sizes.}
    \label{fig:KS}
\end{figure*}

After applying the segmentation mask module (see Section~\ref{sec:segmentation}), \elephant\ finds 1669 hostless candidates, 181 of these have an associated spectroscopic classification available on TNS. Fig.~\ref{fig:TNSclass} compares the TNS classification (horizontal axis) against the classes found on SIMBAD (\texttt{SN*\_candidate}, \texttt{GinCl}, \texttt{SN} and \texttt{Unknown}) or inferred via \fink\ classifiers (Microlensing candidate and SN candidate). We can see that most of these hostless candidates were classified as SN candidate by the ML classifiers, which is consistent with the final spectroscopic classification available on TNS. 

The stamps associated with the 181 hostless candidates with a TNS classification were visually inspected using the Aladin sky atlas\footnote{\url{https://aladin.cds.unistra.fr/}} \citep{2000A&AS..143...33B}. We were not able to visually identify a host for 118 candidates, thus we confirm them as hostless candidates. The remaining 63 events are considered to be contaminants. 
Figure \ref{fig:KS} shows the distribution of the K-S statistic, $D$, for the three considered cropped cutout sizes (see Section~\ref{sec:Fourier}), for both classes, confirmed hostless candidates and  contaminants with host. We used the distribution of the hostless candidates to empirically define a threshold that would enclose a minimum of 75\% of the hostless events. Table \ref{tab:threshold} shows the 75th percentile for each image size. 
Aiming at a low contamination level with 75\% completeness, we chose to use the 15 $\times$  15 pixel images and imposed a threshold of K-S statistic $D <$ 0.5. Thus, we classify all objects with a K-S statistic below the threshold in either the \textit{template} or the \textit{science} image as hostless candidates. 
The last column of Table~\ref{tab:threshold} shows the resulting contamination when the threshold is applied. We note that the output of \elephant\ is the $D$ value, and the user could employ a different threshold to select hostless candidates. In particular, the threshold could be further adjusted once more events are confirmed to be hostless.

\begin{table}
    \centering
    \begin{tabular}{l|c|c}
    Image size (pix) & K-S threshold & Contamination (\%)  \Bstrut \\
    \hline
    \Tstrut
    7 $\times$ 7  &  0.25 & 27.01 \\
    15 $\times$ 15 & 0.50 & 25.97\\
    29 $\times$ 29 & 0.90 & 27.33
    \end{tabular}
    \caption{Kolmogorov-Smirnov statistic thresholds and corresponding contamination levels for different cutout sizes. The threshold was determined using only visually confirmed hostless objects with TNS classification and requiring completeness of 75\%.}
    \label{tab:threshold}
\end{table}

\subsection{Hostless sources on TNS}
\label{subsec:objs}

After applying the K-S $D$ statistic threshold to all the events flagged as hostless candidates by the image segmentation method, we find a total of 1563 ZTF events that match our criteria to be considered hostless candidates. We note that these events are flagged as hostless candidates because no extended source is found at the position of the transient at the center of the stamp. However, the transient could still be associated with a host that is either significantly off-center or that is dimmer than the limiting magnitude of the survey, which for ZTF is $\sim$ 20.5~mag \citep{2019PASP..131a8002B}. To define an event as truly hostless, user inspection is required. The retrieved number of hostless candidates represents $\lesssim$ 2\% of the analyzed extragalactic transients and $\lesssim$ 0.01\% of the number of transients processed by Fink between January/2022 and December/2023.

Among the hostless candidates retrieved after applying the K-S statistic threshold (Section \ref{subsec:threshold}), 154 have an associated spectroscopic classification available on TNS. As the threshold was applied to the complete sample produced by the sigma clipping procedure (Section \ref{sec:segmentation}), including those events for which a host was spotted via visual inspection (see Section\ref{subsec:threshold}), 40 of the 154 hostless candidates with a TNS classification actually have a host that can be identified visually. In other words, the TNS classified hostless candidates present a $\sim$ 26\% contamination, which is consistent with the value reported in Table~\ref{tab:threshold}. 
Table~\ref{tab:TNSsample-spec} the 154 events together with the reported classification. We can see that the most common class is Type Ia SNe, encompassing $\sim$ 67.5\% of events (considering all Type Ia subclasses). This is twice what was found by \cite{2015MNRAS.448.1206M}, but it is consistent with SNe~Ia being predominant among hostless transients. The second most common class is SLSNe, which encompasses $\sim$ 14\% of the sample (considering both SLSNe~I and SLSNe~II). This is also consistent with the results of \cite{2015MNRAS.448.1206M}. In a few cases, a transient reported to TNS is associated to more than one ZTF identifier, Table~\ref{tab:TNSsample-spec} lists all of them, even if they are duplicated, this is because \elephant\ only considers stamps associated to alerts, ignoring the associated coordinates. Inspecting the reasons for the duplicated ZTF identification is out of the scope of this paper.

The last column of Table~\ref{tab:TNSsample-spec} includes comments on some of the events. In particular, we see that a potential, usually faint, host has been reported on TNS for 11 events that we consider to be hostless.
This is compatible with the contamination factor that we report above -- further analysis is needed to confirm these associations. We also notice that eight of our hostless candidates 
were selected by the FLEET \citep[``Finding Luminous and Exotic Extragalactic Transients''][]{Gomez_2020, Gomez_2023} pipeline as potentially luminous or exotic transients. In addition, five of our hostless candidates are part of the sample paper presented by \cite{Chen_23}, that analyzes the characteristics of 78 SLSNe~I. Moreover, three of the SNe reported in Table~\ref{tab:TNSsample-spec} were found in real-time by different groups, followed up, and studied in great detail due to their rare or anomalous nature. Below we provide further details on each of these events.

\begin{figure}
\includegraphics[width=\columnwidth]{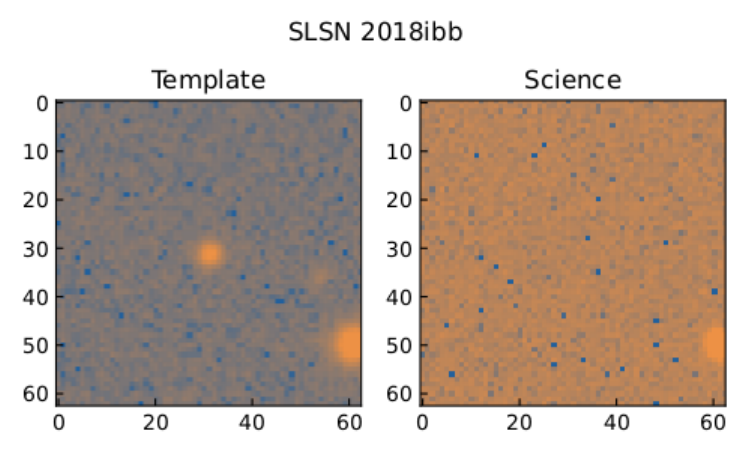}
\caption{Stacked \textit{template} (left) and \textit{science} (right) stamps for SLSN2018ibb (\texttt{ZTF18acenqto}, \texttt{ZTF18adovhai}).}
\label{fig:SN2018ibb}
\end{figure}

\begin{figure}
    \centering
    \includegraphics[width=\columnwidth]{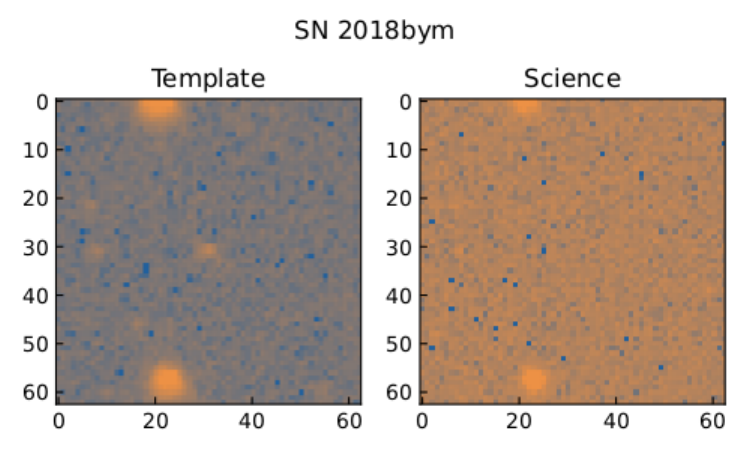}
    \caption{Stacked \textit{template} (left) and \textit{science} (right) stamps for SN2018bym (\texttt{ZTF18aapgrxo}).}
    \label{fig:SN2018bym}
\end{figure}

\begin{figure}
    \centering
    \includegraphics[width=\columnwidth]{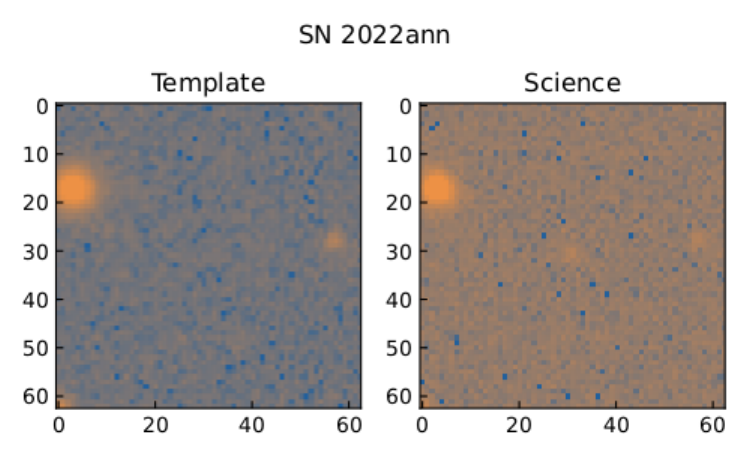}
    \caption{Stacked \textit{template} (left) and \textit{science} (right) stamps for SN2022ann (\texttt{ZTF22aaaihet}).}
    \label{fig:SN2022ann}
\end{figure}

\begin{itemize}
    \item[\Huge{\textbf{.}}] SLSN~2018ibb was identified by \cite{Schulze_2024} as the best pair-instability supernova (PISNe) candidate to date. It has been proposed that PISNe occur when instabilities produced by pair-production induce the thermonuclear explosion of the most massive stars (140~M$_{\sun} < $ M $ < $ 260~M$_{\sun}$). Thus, it has been proposed that PISNe mark the explosive death of Population III stars, which could be indirectly studied through the characteristics of the observed explosion \citep[e.g.][]{2011ApJ...734..102K,2012IAUS..279..253G}. Although SLSN~2018ibb is not hostless, it is associated with a faint \citep[m$_{R}\sim$ 24.4 mag][]{Schulze_2024} dwarf host, detected on 4- and 8-m class telescopes. Thus, for the purposes of the ZTF alerts processed by our pipeline, the transient is expected to appear hostless. Figure \ref{fig:SN2018ibb} illustrates the interesting aspect which lead this object to be detected by our pipeline. It is a typical case of contaminated \textit{template}, meaning that the \textit{template} image was taken when the transient was bright, which results in a relatively lower central brightness in the \textit{science} image. This result demonstrates the importance of considering both sets of stamps in parallel before a decision is made. \\
    \item[\Huge{\textbf{.}}] SN~2018bym was studied by \cite{Lunnan_2020} alongside three other SLSNe discovered by ZTF to examine the origin and diversity of these events. The authors find that SN~2018bym can be considered a classical SLSN~I, and that it is associated with a faint (m$_{r}\sim$ 22.4 mag) dwarf galaxy, for which they obtained deeper observations with the Canada–France–Hawaii Telescope (CFHT). This event is also a representative case of \textit{template} contamination, where the hostless stamp is the \textit{science} one (Figure~\ref{fig:SN2018bym}).
    \item[\Huge{\textbf{.}}] SN~2022ann was studied by \cite{Davis_2023} as one of only five known SNe~Icn. The early discovery of SN~2022ann enabled a detailed analysis of the progenitors of these rare objects. The authors find that SN~2022ann is associated with a faint dwarf host galaxy located in the lower end of the SN host galaxy luminosity distribution. Its  stacked \textit{template} and \textit{science} stamps are shown in Figure~\ref{fig:SN2022ann}.
\end{itemize}

The fact that \elephant\ was able to identify such interesting sources while analyzing historical data demonstrates its potential in identifying similarly interesting objects when applied to more recent alerts. We are currently working on such an investigation and, in parallel, integrating \elephant\ to Fink. Preliminary results are encouraging and will be reported in a subsequent work.  
We also anticipate that, among other applications, the pipeline can serve as a powerful tool to identify SNe potentially associated to dwarf host galaxies \citep[e.g.][]{2021MNRAS.503.3931T}.

\begin{figure}
	\includegraphics[width=\columnwidth]{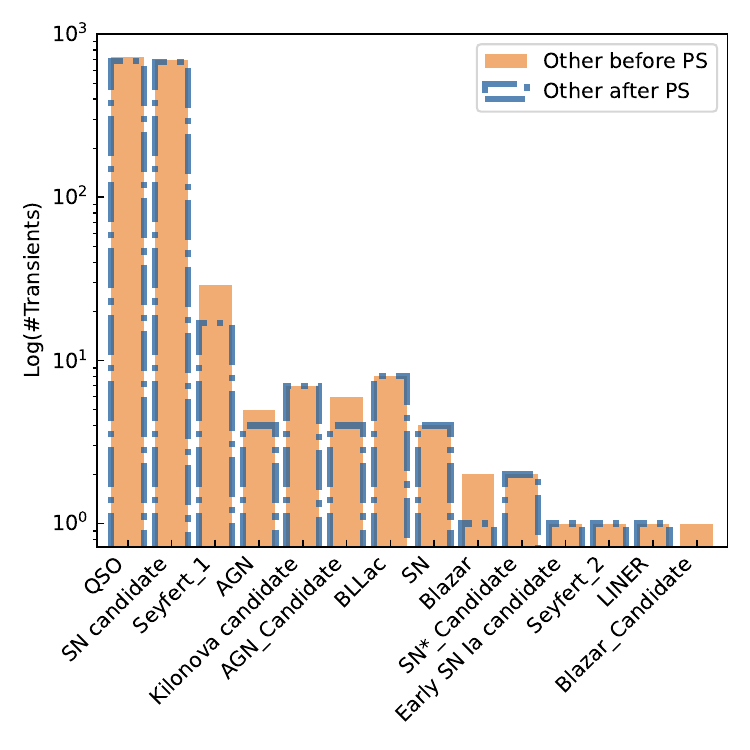}
    \caption{Number of transients without a reported classification on TNS considered to be hostless by the sigma clipping method before applying the power spectrum (PS) analysis (orange), and the number of surviving hostless candidates after applying the PS analysis (blue).}
    \label{fig:MLclass}
\end{figure}

The classification distribution of the hostless candidates that do not have a class available on TNS is shown in Fig.~\ref{fig:MLclass}. The classification associated with these events is mainly obtained from cross-match with SIMBAD or inferred using a ML classifier. We find that $\sim$ 49\% of these events are QSOs, which belong to the family of AGN and, thus, would be associated with a host by definition. However, hostless QSOs have been found before \citep[e.g.][]{2005Natur.437..381M,2010PASP..122..683K}. Although many of the QSOs in our sample of hostless candidates may be associated with a faint, undetected host, \elephant\ can be used to perform systematic searches of hostless QSOs. The other dominant class in this sample is ``SN candidate'',  $\sim$ 48\% of the sample is associated with this ML-based classification. As mentioned above, \elephant\ only considers ZTF stamps associated to individual alerts, however, some events seems to be associated to multiple ZTF identifiers, when this occurs we considered all of the different identifications to be hostless candidates. An interesting case is that of AT~2024dum, this object was found to be a hostless candidate and is associated to three ZTF alerts: ZTF23aabtyzn, ZTF23aaiyhen and ZTF23abkiray. AT~2024dum has been reported to be a fast-moving star (see report\footnote{\url{https://www.wis-tns.org/object/2024dum}} by Shumkov et al. 2024), which could explain the multiplicity of ZTF identifiers. 

\section{Conclusions}
\label{sec:conc}

We developed the \elephant\ pipeline, which processes stamps delivered by the ZTF alert stream and automatically detects hostless transient candidates. The pipeline (see Section~\ref{sec:pipe}) returns stacked \textit{science} and \textit{template} cutouts together with the number of stamps that were used to produce them, a boolean that indicates whether the transient is a hostless candidate based on the segmentation mask analysis (see Section~\ref{sec:segmentation}), the distance to the closest mask in pixels and the associated K-S $D$ statistic obtained from  $7 \times 7$, $15 \times 15$ and $29 \times 29$ pixel square sub-cutouts (see Section~\ref{sec:Fourier}). 

In this work, we define a threshold on the K-S $D$ statistic that is used to flag a transient as a hostless candidate (see Section~\ref{sec:res}). However, future users can use the output values to implement different selection cuts specific to their science case.
The automatic detection of hostless transients has many potential applications that include but are not limited to:
\begin{enumerate}
    \item Identification of transients associated with dwarf and/or dim galaxies to study their characteristics and environments;
    \item Identification of AGNs associated with low-mass galaxies to study their impact on galaxy evolution;
    \item Search of sources that have been ejected from their host galaxies to study intra-cluster stellar populations;
    \item Selection of SNe~Ia and/or SLSNe, as they seem to be predominant among the hostless candidates that have a reported class on TNS;
    \item Compilation of hostless candidates as training sets to improve ML classifiers.
\end{enumerate}

\elephant\ will be included in the Fink broker to allow the real-time detection of hostless candidates and also the retrieval of archival potentially hostless events. Recently, \cite{2024MNRAS.tmp..936Q} performed an statistical analysis of the environments of 161 hostless SNe reported to TNS since 2016. 
They find that their sample is dominated by SNe~Ia and SLSNe, which is in agreement with our findings.  \elephant\ is a useful tool to gather hostless events for similar statistical environmental analyses of different types of SNe. In addition, it can be used to systematically select hostless candidates for classification to increase the number of spectroscopically classified hostless transients to be considered in future population analyses. 

The methods we use here are completely transferable to any dataset by scaling the sizes of the considered stamps. In particular, once the Fink broker starts ingesting LSST alerts, we could test and tune \elephant\ as a tool for finding hostless candidates within the LSST alert stream. LSST is an 8-m class telescope that will have a limiting magnitude of $\sim$25~mag in optical bands \citep{2009arXiv0912.0201L}, which makes it much deeper than the current wide-field surveys. Thus, a big fraction of the events that we flag as hostless candidates here may have a visible host in the LSST stamps. Consequently, if an LSST stamp is flagged to be a hostless by \elephant\ the chances are that the transient is either part of the intra-cluster medium or, it is associated to hosts dimmer than any detected so far. We can only speculate that the hostless transients detected by LSST will be extraordinarily anomalous providing unprecedented insights to the transient sky, with the study of their environments only being possible by using other 8-m class telescopes or by the next generation of large telescopes such as the Giant Magellan Telescope \citep{2012SPIE.8444E..1HJ} and the Extremely Large Telescope \citep{2007Msngr.127...11G}. In this context, automatic pipelines tailored for specific science cases, such as \elephant\,
will play a central role in the process of transient characterization and optimization of follow-up resources. 
\begin{ThreePartTable}
\begin{TableNotes}
      \item First column presents the IAU name of each object. The second column shows the corresponding ZTF internal name. The third and fourth columns show right ascension and declination, respectively. Column five shows the classification available on TNS. In the sixth column we indicate whether we can visually confirm the lack of an obvious host associated with the transient. In the last column we add additional remarks about certain events.
\end{TableNotes}
\onecolumn
\setlength{\tabcolsep}{2pt} % Default value: 6pt
\begin{longtable}{@{\makebox[3em][r]{\rownumber\space}}|lp{2.5cm}p{2cm}p{2cm}lcp{4cm}}
    \caption{Hostless candidates with associated spectroscopic classification in TNS.}
    \label{tab:TNSsample-spec}
    \endfirsthead
    \hline\hline          % inserts double horizontal lines
 IAU Name & ZTF Name  & R.A.     &  Dec.       & Class   &  Confirmed  & 	Comments\\
          &           & [J2000]  &  [J2000]    &         &              &   \gdef\rownumber{\stepcounter{magicrownumbers}\arabic{magicrownumbers}}\\
\hline  
SN~2016ieq    &  \href{https://fink-portal.org/ZTF19abkaxlf}{ZTF19abkaxlf} & 21:22:25.18 &  -11:56:54.82 &          SNIIn   &    $\times$    &  \\
SN~2017iuu    &  \href{https://fink-portal.org/ZTF18aajwbhh}{ZTF18aajwbhh} & 06:27:40.06 &   47:29:45.51 &           SNIa   &    \checkmark  &
Several potential hosts, no redshift info. \\
SN~2018fd     &  \href{https://fink-portal.org/ZTF18adoeywv}{ZTF18adoeywv} & 09:10:36.36 &   35:43:18.39 &         SLSN-I   &    $\times$    &  \\
SN~2018gj     &  \href{https://fink-portal.org/ZTF18aaxljll}{ZTF18aaxljll} & 16:32:02.27 &   78:12:40.96 &           SNII   &    $\times$    &  \\
SN~2018hh     &  \href{https://fink-portal.org/ZTF18aaajfsd}{ZTF18aaajfsd} & 12:13:41.40 &   28:26:39.92 &           SNIa   &    $\times$    &  \\
SN~2018kl     &  \href{https://fink-portal.org/ZTF18aaacdnd}{ZTF18aaacdnd} & 09:09:37.99 &   48:39:39.95 &           SNIa   &    \checkmark  & Potential host association on TNS. \\
SN~2018mc     &  \href{https://fink-portal.org/ZTF18aatpnrf}{ZTF18aatpnrf} & 18:01:00.89 &   61:41:46.76 &          SNIIb   &    $\times$    &  \\
SN~2018vx     &  \href{https://fink-portal.org/ZTF18adkgxye}{ZTF18adkgxye} & 14:43:10.45 &   17:28:16.76 &  SNIa-91T-like  &    $\times$    & Potential host association on TNS. \\
SN~2018vx     &  \href{https://fink-portal.org/ZTF18aaznlwl}{ZTF18aaznlwl} & 14:43:10.44 &   17:28:16.88 &  SNIa-91T-like   &    $\times$  &  \\
SN~2018yc     &  \href{https://fink-portal.org/ZTF18aabqgnb}{ZTF18aabqgnb} & 11:52:45.48 &   37:51:15.44 &           SNIa   &    $\times$    &  \\
SN~2018aae    &  \href{https://fink-portal.org/ZTF18aaiscil}{ZTF18aaiscil} & 12:21:34.21 &   55:34:27.98 &           SNIa   &    \checkmark  & Faint host in Gaia. \\
SN~2018bym    &  \href{https://fink-portal.org/ZTF18aapgrxo}{ZTF18aapgrxo} & 18:43:13.41 &   45:12:28.23 &         SLSN-I   &    \checkmark  & \cite{Lunnan_2020,Chen_23}. \\
SN~2018cog    &  \href{https://fink-portal.org/ZTF18aaxtcdm}{ZTF18aaxtcdm} & 15:26:11.95 &   06:21:25.87 &           SNIa   &    \checkmark  &  \\
SN~2018cxa    &  \href{https://fink-portal.org/ZTF18abfylqx}{ZTF18abfylqx} & 22:28:34.59 &   11:37:05.55 &         SLSN-I   &    $\times$    &  \\
SN~2018eem    &  \href{https://fink-portal.org/ZTF18absoghh}{ZTF18absoghh} & 23:36:01.41 &   18:41:07.06 &           SNII   &    $\times$    &  \\
SN~2018fcg    &  \href{https://fink-portal.org/ZTF18admasii}{ZTF18admasii} & 21:09:36.77 &   33:28:59.43 &         SLSN-I   &    $\times$    &  \\
SN~2018fer    &  \href{https://fink-portal.org/ZTF18abtvstb}{ZTF18abtvstb} & 20:33:05.24 &  -20:51:24.43 &          SNIIb   &    \checkmark  &  \\
SN~2018ffj    &  \href{https://fink-portal.org/ZTF18abslpjy}{ZTF18abslpjy} & 02:30:59.80 &  -17:20:26.84 &         SLSN-I   &    \checkmark  & \cite{GarciaZamora_2018}. \\
SN~2018ftd    &  \href{https://fink-portal.org/ZTF18abotdef}{ZTF18abotdef} & 02:01:16.09 &  -01:13:26.91 &           SNIa   &    \checkmark  &  \\
SN~2018fus    &  \href{https://fink-portal.org/ZTF18abskoyh}{ZTF18abskoyh} & 21:02:31.29 &  -05:37:30.08 &           SNII   &    $\times$    &  \\
SN~2018gck   &  \href{https://fink-portal.org/ZTF18abskzjm}{ZTF18abskzjm} &   00:50:56.6 & 03:29:55.20 &           SNIa   &    \multirow{2}{*}{\checkmark}  &  \\
 SN~2018gck&  \href{https://fink-portal.org/ZTF18adnfkzf}{ZTF18adnfkzf} & 00:50:56.61 &   03:29:55.00 &           SNIa   &      &  \\
SN~2018gkz    &  \href{https://fink-portal.org/ZTF18abvgjyl}{ZTF18abvgjyl} & 07:58:11.55 &   19:31:07.99 &         SLSN-I   &    $\times$    &  \cite{Chen_23}.\\
SN~2018htb    &  \href{https://fink-portal.org/ZTF18acdqmxr}{ZTF18acdqmxr} & 04:37:30.67 &   20:16:55.70 &           SNIa   &    $\times$    &  \\
SN~2018ibb    &  \href{https://fink-portal.org/ZTF18acenqto}{ZTF18acenqto} & 04:38:56.94 &  -20:39:44.06 &         SLSN-I   &    \checkmark  & \cite{Schulze_2024,Chen_23}. \\
SN~2018ibb    &  \href{https://fink-portal.org/ZTF18adovhai}{ZTF18adovhai} & 04:38:56.95 &  -20:39:43.93 &         SLSN-I   &    \checkmark  &  \\
SN~2018icz    &  \href{https://fink-portal.org/ZTF18accngfb}{ZTF18accngfb} & 10:03:14.82 &   15:04:42.87 &           SNIa   &    \checkmark  & Gaia hostless candidate. \\
SN~2018imd    &  \href{https://fink-portal.org/ZTF18acydvjn}{ZTF18acydvjn} & 12:48:24.97 &  -05:47:39.10 &           SNIa   &    $\times$    &  \\
SN~2018imq    &  \href{https://fink-portal.org/ZTF18acepwhb}{ZTF18acepwhb} & 11:34:45.61 &   77:03:09.99 &           SNIa   &    \checkmark  &  \\
SN~2018jeo    &  \href{https://fink-portal.org/ZTF18aczddnw}{ZTF18aczddnw} & 09:04:36.91 &  -19:47:09.60 &           SNIa   &    $\times$    &   \\
SN~2018lzw    &  \href{https://fink-portal.org/ZTF18abrzcbp}{ZTF18abrzcbp} & 07:39:32.76 &   27:44:02.62 &         SLSN-I   &    \checkmark  &  \cite{Chen_23}.\\
SN~2018lzx    &  \href{https://fink-portal.org/ZTF18abszecm}{ZTF18abszecm} & 22:29:27.23 &   13:10:39.96 &         SLSN-I   &    \checkmark  &  \cite{Chen_23}.\\
SN~2019aatt   &  \href{https://fink-portal.org/ZTF19abszdld}{ZTF19abszdld} & 01:21:21.63 &   30:17:03.52 &           SNIa   &     $\times$   & \\		   
SN~2020jhs    &  \href{https://fink-portal.org/ZTF20aayvmyh}{ZTF20aayvmyh} & 09:28:14.10 &   25:40:13.39 &          SNIIn   &    \checkmark  &  \\
SN~2021rll    &  \href{https://fink-portal.org/ZTF21abiwpjm}{ZTF21abiwpjm} & 13:45:21.99 &   26:45:00.72 &          SNIIn   &    \checkmark  &  Faint host in Pan-STARRS. \\
SN~2022aj     &  \href{https://fink-portal.org/ZTF22aaafohf}{ZTF22aaafohf} & 14:56:08.32 &  -27:45:37.53 &           SNIa   &    \checkmark  & Gaia hostless candidate. \\
SN~2022aj     &  \href{https://fink-portal.org/ZTF22aaausrb}{ZTF22aaausrb} & 14:56:08.31 &  -27:45:37.58 &           SNIa   &    \checkmark  & Gaia hostless candidate.\\
SN~2022ait    &  \href{https://fink-portal.org/ZTF22aaaiykj}{ZTF22aaaiykj} & 10:30:26.97 &   07:10:21.19 &           SNIa   &    \checkmark  &  \\
SN~2022ann    &  \href{https://fink-portal.org/ZTF22aaaihet}{ZTF22aaaihet} & 10:17:29.66 &  -02:25:35.40 &          SNIcn   &    \checkmark  & \cite{Davis_2023}. \\
SN~2022are    &  \href{https://fink-portal.org/ZTF22aaahull}{ZTF22aaahull} & 09:59:07.08 &  -18:11:02.83 &           SNIa   &    \checkmark  &  \\
SN~2022bic    &  \href{https://fink-portal.org/ZTF22aaagvyp}{ZTF22aaagvyp} & 08:39:08.93 &   60:59:16.25 &           SNIa   &    \checkmark  &  \\
SN~2022cjv    &  \href{https://fink-portal.org/ZTF22aaafavg}{ZTF22aaafavg} & 11:34:34.66 &   31:02:40.71 &           SNIa   &    \checkmark  &  \\
SN~2022ddh    &  \href{https://fink-portal.org/ZTF22aabtyli}{ZTF22aabtyli} & 10:28:15.83 &   06:34:47.31 &           SNIa   &    $\times$    &  \\
SN~2022dld    &  \href{https://fink-portal.org/ZTF22aabwvot}{ZTF22aabwvot} & 14:06:16.60 &   13:29:30.89 &           SNIa   &    \checkmark  & FLEET Candidate. \\
SN~2022fjx    &  \href{https://fink-portal.org/ZTF22aadlmgg}{ZTF22aadlmgg} & 10:43:30.16 &   19:04:58.70 & SNIa-91bg-like   &    \checkmark  &  \\
SN~2022ful    &  \href{https://fink-portal.org/ZTF22aadeuwu}{ZTF22aadeuwu} & 19:20:10.68 &   50:23:42.41 &         SLSN-I   &	 \checkmark  & Gaia hostless candidate. \\
SN~2022ful    &  \href{https://fink-portal.org/ZTF22aafumyr}{ZTF22aafumyr} & 19:20:10.67 &   50:23:42.40 &         SLSN-I   &    \checkmark  & Gaia hostless candidate. \\
SN~2022gkv    &  \href{https://fink-portal.org/ZTF22aaftcmp}{ZTF22aaftcmp} & 15:57:51.12 &   29:55:10.78 &           SNIa   &    \checkmark  &  \\
SN~2022gkv    &  \href{https://fink-portal.org/ZTF22aadetzs}{ZTF22aadetzs} & 15:57:51.12 &   29:55:10.82 &           SNIa   &    \checkmark  &  \\
SN~2022gsp    &  \href{https://fink-portal.org/ZTF22aadqkgp}{ZTF22aadqkgp} & 14:53:08.28 &   13:59:57.53 &           SNIa   &    $\times$    &  \\
SN~2022hdn    &  \href{https://fink-portal.org/ZTF22aagbxrb}{ZTF22aagbxrb} & 15:00:09.14 &   36:07:13.14 &        SNIc-BL   &    \checkmark  & Potential host association on TNS. \\
SN~2022huk    &  \href{https://fink-portal.org/ZTF22aahaasc}{ZTF22aahaasc} & 10:14:12.85 &  -23:41:17.10 &           SNIa   &	 \checkmark  & \\
SN~2022hwk    &  \href{https://fink-portal.org/ZTF22aagzbux}{ZTF22aagzbux} & 12:45:59.22 &   59:15:37.04 &          SNIIn   &    $\times$    &  \\
SN~2022igq    &  \href{https://fink-portal.org/ZTF22aahecwj}{ZTF22aahecwj} & 13:56:52.02 &   19:07:01.66 &           SNIa   &    \checkmark  &  \\
SN~2022ihz    &  \href{https://fink-portal.org/ZTF22aahgxdt}{ZTF22aahgxdt} & 09:42:48.23 &  -03:36:25.48 &  SNIa-91bg-like  &   \checkmark   & \\
SN~2022irt    &  \href{https://fink-portal.org/ZTF22aahhubz}{ZTF22aahhubz} & 12:27:12.57 &   00:55:40.00 &           SNIa   &    \checkmark  &  \\
SN~2022jii    &  \href{https://fink-portal.org/ZTF22aaizxqg}{ZTF22aaizxqg} & 14:54:31.30 &   04:19:52.83 &           SNIa   &    \checkmark  & Potential host association on TNS. \\
SN~2022jnr    &  \href{https://fink-portal.org/ZTF22aajhtpy}{ZTF22aajhtpy} & 15:02:39.48 &   17:14:23.45 &           SNIa   &    $\times$    &  \\
SN~2022jzt    &  \href{https://fink-portal.org/ZTF22aakanzk}{ZTF22aakanzk} & 13:43:12.79 &   48:23:10.82 &           SNIa   &    \checkmark  & Potential host association on TNS. \\
SN~2022knm    &  \href{https://fink-portal.org/ZTF22aakkmri}{ZTF22aakkmri} & 13:25:04.36 &  -24:39:24.94 &           SNIa   &    \checkmark  &  \\
SN~2022llq    &  \href{https://fink-portal.org/ZTF22aalmrqp}{ZTF22aalmrqp} & 12:03:16.55 &   51:49:54.24 &           SNIa   &    \checkmark  &  \\
SN~2022lxd    &  \href{https://fink-portal.org/ZTF22aaljlzq}{ZTF22aaljlzq} & 17:36:38.67 &   61:33:18.66 &         SLSN-I   &    \checkmark  &  \\
SN~2022mjk    &  \href{https://fink-portal.org/ZTF22aapuake}{ZTF22aapuake} & 01:25:41.36 &   01:45:41.27 &           SNIa   &    \checkmark  &  \\
SN~2022nab    &  \href{https://fink-portal.org/ZTF22aaobrbd}{ZTF22aaobrbd} & 18:38:57.89 &   48:23:04.86 &           SNIa   &    \checkmark  &  \\
AT~2022nci    &  \href{https://fink-portal.org/ZTF22aaombjf}{ZTF22aaombjf} & 00:46:33.41 &   41:45:35.15 &           Nova   &    \checkmark  &  \\
SN~2022ncx    &  \href{https://fink-portal.org/ZTF22aaogwbd}{ZTF22aaogwbd} & 12:08:13.50 &   66:38:24.84 &  SNIa-91T-like   &    \checkmark  &  \\
SN~2022ojm    &  \href{https://fink-portal.org/ZTF22aapjqpn}{ZTF22aapjqpn} & 23:37:46.03 &   40:05:07.96 &         SLSN-I   &    \checkmark  &  \\
SN~2022orr    &  \href{https://fink-portal.org/ZTF22aasaapb}{ZTF22aasaapb} & 15:50:58.27 &   68:35:07.80 &           SNIa   &    \checkmark  &  \\
SN~2022owf    &  \href{https://fink-portal.org/ZTF22aaszlph}{ZTF22aaszlph} & 23:26:09.97 &   27:42:02.97 &           SNIa   &    \checkmark  &  \\
SN~2022rfn    &  \href{https://fink-portal.org/ZTF22abahblc}{ZTF22abahblc} & 19:11:28.98 &  -17:11:07.59 &           SNIa   &    $\times$    &  \\
SN~2022rhl    &  \href{https://fink-portal.org/ZTF22aasoali}{ZTF22aasoali} & 19:20:44.21 &   46:52:54.75 &          SNIIn   &    \checkmark  &  \\
SN~2022rpm    &  \href{https://fink-portal.org/ZTF22abamxcl}{ZTF22abamxcl} & 02:01:11.36 &  -05:51:59.41 &             SN   &    \checkmark  & \\
SN~2022sff    &  \href{https://fink-portal.org/ZTF22abdibiz}{ZTF22abdibiz} & 07:56:05.03 &   33:28:18.38 &           SNIa   &    $\times$    &  \\
SN~2022tis    &  \href{https://fink-portal.org/ZTF22abepfmn}{ZTF22abepfmn} & 21:10:35.86 &  -09:30:14.39 &           SNII   &    \checkmark  &  \\
SN~2022uhk    &  \href{https://fink-portal.org/ZTF22abfwchw}{ZTF22abfwchw} & 18:50:17.25 &   75:27:59.88 &           SNII   &    \checkmark  &  \\
SN~2022uot    &  \href{https://fink-portal.org/ZTF22abfyvhf}{ZTF22abfyvhf} & 05:37:10.51 &   68:34:31.96 &          SNIIn   &    \checkmark  &  \\
SN~2022uwh    &  \href{https://fink-portal.org/ZTF22abfxmvf}{ZTF22abfxmvf} & 23:53:37.16 &   11:22:58.08 &           SNIa   &    \checkmark  &  \\
SN~2022wlm    &  \href{https://fink-portal.org/ZTF22abjafpr}{ZTF22abjafpr} & 05:56:46.63 &   48:06:20.85 &        SNIc-BL   &    \checkmark  &  \\
SN~2022wpp    &  \href{https://fink-portal.org/ZTF22abjrpmv}{ZTF22abjrpmv} & 16:41:49.91 &   15:15:45.35 &           SNIa   &    \checkmark  &  \\
SN~2022wuw    &  \href{https://fink-portal.org/ZTF22ablcybb}{ZTF22ablcybb} & 16:26:19.28 &   80:28:41.33 &           SNIa   &    \checkmark  &  \\
SN~2022wuy    &  \href{https://fink-portal.org/ZTF22ablhldn}{ZTF22ablhldn} & 06:44:23.34 &   32:14:53.21 &           SNIa   &    \checkmark  &  \\
SN~2022xjl    &  \href{https://fink-portal.org/ZTF22abmpqbq}{ZTF22abmpqbq} & 23:57:11.78 &   05:36:17.35 &           SNIa   &    \checkmark  &  \\
SN~2022xxn    &  \href{https://fink-portal.org/ZTF22abmxtqr}{ZTF22abmxtqr} & 01:18:56.59 &  -12:57:44.93 &           SNIa   &    \checkmark  &  \\
SN~2022ycr    &  \href{https://fink-portal.org/ZTF22abnwvyc}{ZTF22abnwvyc} & 21:23:27.18 &  -18:06:13.85 &          Other   &    $\times$    &  \\
SN~2022ydl    &  \href{https://fink-portal.org/ZTF22abnqzle}{ZTF22abnqzle} & 22:40:04.43 &  -06:38:28.35 &           SNIa   &    \checkmark  &  \\
SN~2022yig    &  \href{https://fink-portal.org/ZTF22aboaiim}{ZTF22aboaiim} & 05:20:21.53 &  -20:54:41.61 &           SNIa   &    $\times$    &  \\
SN~2022yru    &  \href{https://fink-portal.org/ZTF22aboixdd}{ZTF22aboixdd} & 10:27:28.41 &   70:59:02.23 &           SNIa   &    $\times$    & \\	   
AT~2022zzj    & \href{https://fink-portal.org/ZTF22abtltcw}{ZTF22abtltcw}  & 00:41:25.73 &   40:44:23.34 &           Nova   &     $\times$   &  Potential host association on TNS.\\
SN~2022aahy   &  \href{https://fink-portal.org/ZTF22abtsypf}{ZTF22abtsypf} & 06:58:56.24 &   39:38:06.90 &          SNIIn   &    $\times$    &  \\
SN~2022aahz   &  \href{https://fink-portal.org/ZTF22abtotgu}{ZTF22abtotgu} & 12:25:54.64 &   06:45:02.96 &           SNIa   &    \checkmark  &  \\
SN~2022abtm   &  \href{https://fink-portal.org/ZTF22abvngdr}{ZTF22abvngdr} & 23:03:54.16 &   15:46:19.84 &         SLSN-I   &    \checkmark  &  \\
SN~2022acfw   & \href{https://fink-portal.org/ZTF22abzakdd}{ZTF22abzakdd}  & 13:21:06.78 &   27:54:53.79 &           SNIa   &     $\times$   & \\	   
SN~2022acmr   &  \href{https://fink-portal.org/ZTF22abyhqkt}{ZTF22abyhqkt} & 02:02:39.45 &  -07:02:22.67 &           SNIa   &    \checkmark  &  \\
SN~2022acsx   &  \href{https://fink-portal.org/ZTF22abynkpz}{ZTF22abynkpz} & 06:12:59.10 &   68:48:45.39 &         SLSN-I   &    \checkmark  & Faint host in DESI Legacy Surveys DR10. \\
SN~2022adbl   &  \href{https://fink-portal.org/ZTF22abyuoan}{ZTF22abyuoan} & 07:57:29.24 &   62:25:39.25 &           SNIa   &    $\times$    &  \\
SN~2022adrs   &  \href{https://fink-portal.org/ZTF22abzbyyw}{ZTF22abzbyyw} & 00:27:08.37 &  -24:53:50.88 &           SNIa   &    \checkmark  &  \\
SN~2022advb   &  \href{https://fink-portal.org/ZTF22abyznto}{ZTF22abyznto} & 09:40:44.48 &   05:10:21.13 &           SNIa   &    \checkmark  &  \\
SN~2022adxq   &  \href{https://fink-portal.org/ZTF22abzvyku}{ZTF22abzvyku} & 03:27:24.99 &  -17:37:50.35 &           SNIa   &    \checkmark  &  \\
SN~2023ha     &  \href{https://fink-portal.org/ZTF23aaajtqn}{ZTF23aaajtqn} & 09:19:32.46 &  -01:11:34.62 &           SNIa   &    $\times$    &  \\
SN~2023ael    &  \href{https://fink-portal.org/ZTF23aaawbsy}{ZTF23aaawbsy} & 17:14:41.46 &   66:51:22.60 &           SNIa   &    \checkmark  & Gaia hostless candidate.	 \\  
SN~2023aiw    &  \href{https://fink-portal.org/ZTF23aaawcvx}{ZTF23aaawcvx} & 16:31:09.06 &   39:47:20.59 &           SNIa   &    $\times$    &  \\
SN~2023ayq    &  \href{https://fink-portal.org/ZTF23aaazegi}{ZTF23aaazegi} & 13:24:05.23 &  -03:33:41.04 &       SNIa-CSM   &    \checkmark  &  \\
SN~2023bee    &  \href{https://fink-portal.org/ZTF23aabtgej}{ZTF23aabtgej} & 08:56:11.63 &  -03:19:32.05 &           SNIa   &    $\times$    &  \\
SN~2023cpq    &  \href{https://fink-portal.org/ZTF23aacdnjz}{ZTF23aacdnjz} & 17:29:20.16 &   14:11:04.51 &       SNIa-CSM   &    \checkmark  & FLEET Candidate. \\
SN~2023cze    &  \href{https://fink-portal.org/ZTF23aadbswn}{ZTF23aadbswn} & 15:05:05.39 &   28:28:52.45 &           SNIa   &    \checkmark  &  \\
SN~2023ebb    &  \href{https://fink-portal.org/ZTF23aadruma}{ZTF23aadruma} & 11:24:34.69 &   46:53:37.14 &           SNII   &    \checkmark  &  \\
SN~2023erb    &  \href{https://fink-portal.org/ZTF23aaejvzv}{ZTF23aaejvzv} & 16:37:54.80 &   43:23:08.60 &           SNIa   &    $\times$    & \\
SN~2023exi    &  \href{https://fink-portal.org/ZTF23aaelzdb}{ZTF23aaelzdb} & 07:04:14.64 &   67:37:32.53 &           SNIa   &    \checkmark  &  \\
SN~2023ffw    &  \href{https://fink-portal.org/ZTF23aaemgto}{ZTF23aaemgto} & 11:35:12.53 &  -13:30:47.90 &           SNIa   &    \checkmark  &  \\
SN~2023fvf    &  \href{https://fink-portal.org/ZTF23aafggjj}{ZTF23aafggjj} & 13:20:51.42 &   15:17:37.91 &           SNIa   &    $\times$    &  \\
SN~2023gav    &  \href{https://fink-portal.org/ZTF23aaftouh}{ZTF23aaftouh} & 10:47:18.35 &  -05:07:22.75 &           SNIa   &    \checkmark  & FLEET Candidate. \\
SN~2023ger    &  \href{https://fink-portal.org/ZTF23aagaiju}{ZTF23aagaiju} & 10:57:13.46 &   42:58:50.06 &           SNIa   &    \checkmark  &  \\
SN~2023ghq    &  \href{https://fink-portal.org/ZTF23aagunkc}{ZTF23aagunkc} & 15:58:13.05 &   08:54:24.99 &          Other   &    \checkmark  &  \\
SN~2023hoz    &  \href{https://fink-portal.org/ZTF23aagdbbv}{ZTF23aagdbbv} & 16:18:21.55 &   01:31:43.41 &         SLSN-I   &    \checkmark  & FLEET Candidate. \\
SN~2023hrn    &  \href{https://fink-portal.org/ZTF23aaiyexs}{ZTF23aaiyexs} & 11:08:35.07 &   04:48:52.09 &           SNIa   &    $\times$    &  \\
SN~2023huv    &  \href{https://fink-portal.org/ZTF23aafmjbx}{ZTF23aafmjbx} & 13:29:03.66 &  -10:25:29.42 &          SNIIn   &    \checkmark  & FLEET Candidate. \\
SN~2023iar    &  \href{https://fink-portal.org/ZTF23aajenxf}{ZTF23aajenxf} & 13:31:36.14 &   04:55:21.32 &           SNIa   &    \checkmark  &  \\
SN~2023ifa    &  \href{https://fink-portal.org/ZTF23aajhtuu}{ZTF23aajhtuu} & 09:33:34.73 &   51:36:54.11 &          SNIbn   &    $\times$    &  \\
SN~2023iwy    &  \href{https://fink-portal.org/ZTF23aakmewi}{ZTF23aakmewi} & 18:00:18.62 &   26:24:32.04 &        SNIc-BL   &    \checkmark  &  \\
SN~2023jsb    &  \href{https://fink-portal.org/ZTF23aamfmqm}{ZTF23aamfmqm} & 21:49:58.44 &   14:09:24.71 &           SNIa   &    \checkmark  &  \\
SN~2023jvu    &  \href{https://fink-portal.org/ZTF23aamqonh}{ZTF23aamqonh} & 16:38:00.64 &   55:24:18.30 &           SNIa   &    $\times$    &  \\
SN~2023khp    &  \href{https://fink-portal.org/ZTF23aamsekn}{ZTF23aamsekn} & 00:17:56.20 &   23:59:03.33 &       SNIa-CSM   &    \checkmark  & \\
SN~2023kkh    &  \href{https://fink-portal.org/ZTF23aanuvih}{ZTF23aanuvih} & 18:49:05.23 &   45:07:10.91 &           SNIa   &    \checkmark  &  \\
SN~2023kki    &  \href{https://fink-portal.org/ZTF23aamxeoe}{ZTF23aamxeoe} & 16:53:18.11 &   37:41:23.52 &          SNIIn   &    \checkmark  &  \\
SN~2023kvk    &  \href{https://fink-portal.org/ZTF23aanukvi}{ZTF23aanukvi} & 20:02:32.44 &  -05:05:16.88 &           SNIa   &    \checkmark  &  \\
SN~2023mhj    &  \href{https://fink-portal.org/ZTF23aapvrkk}{ZTF23aapvrkk} & 00:00:42.10 &  -12:14:12.51 &           SNIa   &    \checkmark  & \\
SN~2023mir    &  \href{https://fink-portal.org/ZTF23aaqfdby}{ZTF23aaqfdby} & 22:40:59.46 &  -05:04:15.74 &           SNIa   &    \checkmark  &  \\
SN~2023mit    &  \href{https://fink-portal.org/ZTF23aaqjuux}{ZTF23aaqjuux} & 16:28:18.62 &   23:10:56.15 &           SNIa   &    \checkmark  &  \\
SN~2023nbf    &  \href{https://fink-portal.org/ZTF23aawcygl}{ZTF23aawcygl} & 13:08:46.93 &   49:24:08.67 &           SNIa   &    $\times$    &  \\
SN~2023pjx    &  \href{https://fink-portal.org/ZTF23aaxyawz}{ZTF23aaxyawz} & 01:31:25.59 &   12:24:31.35 &           SNIa   &    \checkmark  &  \\
SN~2023qpe    &  \href{https://fink-portal.org/ZTF23aazrtdy}{ZTF23aazrtdy} & 23:40:15.79 &   15:27:50.97 &           SNIa   &    \checkmark  & Potential host association on TNS. \\
SN~2023qrz    &  \href{https://fink-portal.org/ZTF23aaznifc}{ZTF23aaznifc} & 21:55:14.72 &  -17:35:31.55 &           SNIa   &    \checkmark  &  \\
SN~2023qvl    &  \href{https://fink-portal.org/ZTF23aawhcjb}{ZTF23aawhcjb} & 15:50:08.30 &   53:39:37.03 &         SLSN-I   &    \checkmark  & FLEET Candidate. \\
SN~2023qzo    &  \href{https://fink-portal.org/ZTF23abaderr}{ZTF23abaderr} & 01:35:05.20 &  -22:40:37.84 &           SNIa   &    $\times$    &  \\
SN~2023rbt    &  \href{https://fink-portal.org/ZTF23abaslfm}{ZTF23abaslfm} & 01:46:31.43 &   11:51:55.34 &           SNIa   &    \checkmark  &  \\
SN~2023rfg    &  \href{https://fink-portal.org/ZTF23abavpyk}{ZTF23abavpyk} & 23:31:06.17 &  -27:00:56.44 &           SNIa   &    \checkmark  &  \\
SN~2023slt    &  \href{https://fink-portal.org/ZTF23abbsfxp}{ZTF23abbsfxp} & 03:34:04.42 &  -21:54:21.92 &           SNIa   &	 \checkmark  & FLEET candidate. \\
SN~2023spg    &  \href{https://fink-portal.org/ZTF23abcufxh}{ZTF23abcufxh} & 00:11:56.30 &  -07:46:19.56 &           SNIa   &    \checkmark  &  \\
SN~2023svf    &  \href{https://fink-portal.org/ZTF23abcqzvm}{ZTF23abcqzvm} & 16:44:05.26 &   30:18:09.99 &           SNIa   &    \checkmark  &  \\
SN~2023syg    &  \href{https://fink-portal.org/ZTF23abdynfn}{ZTF23abdynfn} & 20:49:40.99 &  -14:43:26.57 &           SNIa   &    \checkmark  &  \\
SN~2023szi    &  \href{https://fink-portal.org/ZTF23aaznlgb}{ZTF23aaznlgb} & 22:19:56.03 &   25:54:56.17 &         SLSN-I   &    \checkmark  & FLEET Candidate. \\
SN~2023tqm    &  \href{https://fink-portal.org/ZTF23abgvtxr}{ZTF23abgvtxr} & 07:57:28.61 &   51:07:23.93 &           SNIa   &     \checkmark & \\
SN~2023upt    &  \href{https://fink-portal.org/ZTF23abjqxbe}{ZTF23abjqxbe} & 04:28:27.38 &  -17:53:27.26 &           SNIa   &    \checkmark  &  \\
SN~2023uqu    &  \href{https://fink-portal.org/ZTF23abijopy}{ZTF23abijopy} & 17:47:47.83 &   64:20:57.31 &           SNIa   &    \checkmark  & Potential host association on TNS. \\
SN~2023vkz    &  \href{https://fink-portal.org/ZTF23ablspnz}{ZTF23ablspnz} & 08:28:36.18 &   57:12:31.86 &           SNII   &    \checkmark  &  \\
SN~2023wml    &  \href{https://fink-portal.org/ZTF23aboebgh}{ZTF23aboebgh} & 11:39:08.69 &  -11:14:57.92 &         SLSN-I   &    \checkmark  &  \\
SN~2023wrn    &  \href{https://fink-portal.org/ZTF23aboemfi}{ZTF23aboemfi} & 23:31:53.52 &   22:39:30.74 &           SNIa   &    \checkmark  &  \\
SN~2023wtq    &  \href{https://fink-portal.org/ZTF23abochfb}{ZTF23abochfb} & 01:21:55.68 &  -03:46:19.95 &        SNIc-BL   &    \checkmark  &  \\
SN~2023xjs    &  \href{https://fink-portal.org/ZTF23abpqklj}{ZTF23abpqklj} & 02:26:34.56 &  -19:10:26.42 &           SNIa   &    \checkmark  & Faint host in Pan-STARRS. \\
SN~2023yqq    &  \href{https://fink-portal.org/ZTF23abryfga}{ZTF23abryfga} & 22:56:04.96 &   19:34:56.71 &           SNIa   &    \checkmark  &  \\
SN~2023yti    &  \href{https://fink-portal.org/ZTF23absflyh}{ZTF23absflyh} & 02:06:22.20 &  -18:19:06.12 &           SNIa   &    \checkmark  &  \\
SN~2023zeq    &  \href{https://fink-portal.org/ZTF23abqygjv}{ZTF23abqygjv} & 01:08:54.20 &  -20:38:24.23 &        SLSN-II   &    \checkmark  &  \\
SN~2023zjv    &  \href{https://fink-portal.org/ZTF23absbyol}{ZTF23absbyol} & 07:14:10.46 &   36:09:52.82 &           SNIa   &    $\times$    &  \\
SN~2023aajn   &  \href{https://fink-portal.org/ZTF23abvbwys}{ZTF23abvbwys} & 03:41:33.85 &  -02:46:50.01 &  SNIa-91T-like   &    \checkmark  &  \\
\hline
\insertTableNotes
\end{longtable}
\twocolumn
\end{ThreePartTable}

\section*{Acknowledgements}

We thank Julien Peloton for assistance in retrieving data from \fink. P.J.P. acknowledges support from the European Research Council (ERC) under the European Union's Horizon Europe research and innovation program (grant agreement No. 10104229 – TransPIre). EEH is supported by a Gates Cambridge Scholarship (\#OPP1144). This work is a result of the COIN Residence Program \#7\footnote{\url{https://cosmostatistics-initiative.org/residence-programs/crp7/}}, held in Lisbon, Portugal, from 9 to 16 September 2023 and supported by the Portuguese Fundação para a Ciência e a Tecnologia (FCT) through the Strategic Programme UIDP/FIS/00099/2020 and UIDB/FIS/00099/2020 for CENTRA. The Cosmostatistics Initiative (COIN, \url{https://cosmostatistics-initiative.org/}) is an international network of researchers whose goal is to foster interdisciplinarity inspired by astronomy. This research has made use of the SIMBAD database, operated at CDS, Strasbourg, France. This research has made use of ``Aladin sky atlas'' developed at CDS, Strasbourg Observatory, France. This work made use of Astropy (\url{http://www.astropy.org}) a community-developed core Python package and an ecosystem of tools and resources for astronomy \citep{astropy:2013, astropy:2018, astropy:2022}.  The \elephant\ icon was taken from \url{https://icons8.com}. The color palette used in this work was  inspired by ``The Temptation of St. Anthony'' by Salvador Dali, 1946. 
%%%%%%%%%%%%%%%%%%%%%%%%%%%%%%%%%%%%%%%%%%%%%%%%%%
\section*{Data Availability}

The data used here can be accessed via the \fink\ data transfer service: \url{https://fink-portal.org/download}. The \elephant\ pipeline is publicly available on \texttt{github}: \url{https://github.com/COINtoolbox/extragalactic_hostless}.



\bibliography{ref}

%%%%%%%%%%%%%%%%% APPENDICES %%%%%%%%%%%%%%%%%%%%%

\appendix

\section{Nearest neighbor}
\label{app:neighbor}

To compute the distance from the transient to the nearest mask, we assume an origin position at the center of the image from where to calculate the distance.
%, i.e. (30,30), in order to calculate distances between the transient and its closest neighbor. 
Masked pixels have value 1, whereas background pixels have value 0. To find the nearest masked pixel to the transient, we compare pixel by pixel until a mask is found. 
First, we check all adjacent pixels (including diagonally adjacent ones) related to the central pixel, starting from left to right and top to bottom. This process is repeated for outer layers until a masked pixel is found or we reach the limits of the 7-pixel threshold. If the distance between the central pixel and the nearest neighbor is not within a 7-pixel radius, we keep that alert in our sample for further checks. %Note that a simple Euclidean distance in pixels does not confirm if a transient is hostless but helps us increase the purity of our hostless candidate selection. We return the Euclidean distance between the nearest masked pixel and the central pixel. 
The Euclidean distance between the nearest masked pixel and the central pixel is included for the user to assess whether a neighbor is close enough for it to be considered an associated host.
The algorithm works as follows:

%\textcolor{red}{BM: Will have a go at writing the pseudocode tomorrow. Have imported a different package for it. (Done.)}
\begin{algorithm}[!htbp]
    \label{alg:find_neighbour}
    %\caption{Pseudocode for nearest-neighbor identification. 
    %\textcolor{red}{BM: Reminder for myself: The template messed the output list up, will fix. (Done.)}
    %}
    \KwData{\hspace{4pt} $\boldsymbol{M}$: Stamp of the segmented image\\
    \hspace{31pt} $r$: Maximum radius for nearest detected source}
    \vspace{8pt}
    \KwResult{$x_1$: x-axis index of the closest masked source\\
    \hspace{30pt} $x_2$: y-axis index of the closest masked source\\
    \hspace{31pt} $\delta$: Euclidean distance measured in image pixels\\
    \hspace{31pt} $f$: Flag (true for hostless, zero otherwise)}
    \vspace{8pt}
    $\boldsymbol{M}_\textup{>0} \leftarrow 1$\\
    $\tau, \delta, f \leftarrow 30, 100, \textup{True}$\\
    $x_1, x_2 \leftarrow \textup{None}$\\
    \vspace{8pt}
    \uIf{$\boldsymbol{M}_{\tau, \tau} = 1$}{$f \leftarrow \textup{False}$\\
    \Return $\tau, \tau, 0, f$}
    \uElse{
    \For{$s \in \{1, \dots , r + 1\}$}{
    $\boldsymbol{a} \leftarrow \{\tau - 1 - s, \tau + 2 + s\}$\\
    $\nu \leftarrow 0$\\
    \For{$i \in \boldsymbol{a}$}{
    \uIf{$(\nu = 0) \lor (\nu = \tau + 1 + s)$}{
    \For{$j \in \textup{a}$}{
    \uIf{$\boldsymbol{M}_{i, j} = 1$}{
    $\delta \leftarrow \sqrt{(\tau - i)^2 + (\tau - j)^2}$\\
    $x_1, x_2, f \leftarrow i, j, \textup{False}$\\
    \Return $x_1, x_2, \delta, f$}}}
    \uElse{
    \For{$j \in \{ \boldsymbol{a}_0, \boldsymbol{a}_{|a|} \}$}{
    \uIf{$\boldsymbol{M}_{i, j} = 1$}{
    $\delta \leftarrow \sqrt{(\tau - i)^2 + (\tau - j)^2}$\\
    $x_1, x_2, f \leftarrow i, j, \textup{False}$\\
    \Return $x_1, x_2, \delta, f$}}}
    $\nu \leftarrow \nu + 1$}}}
    \Return $x_1, x_2, \delta, f$
    \vspace{10pt}
\end{algorithm}

\iffalse
\begin{algorithm}
  \caption{Finding the nearest neighbour}
  \label{euclid}
  \begin{algorithmic}[1]
    \Require $\text{matrix}[i_T][i_T]==0$ \Comment{Transient is outside the contours of a detection}
    \Function{calculate\_distance}{$matrix, radius$}
      \While{$step < N-1$ }
      \State $array \gets [i_{T} -1 -step, ..., i_{T}+1+step]$
        \For{each i in array}
        % \if{$index==0 or index==i_{c}+1+step$} 
      % \Endif  
      \Endfor
      \label{euclidendwhile}
      \State \Return{$i, j, \text{distance}$}
    \EndFunction
  \end{algorithmic}
\end{algorithm}
\fi

\section{Segmentation masks with \texttt{SExtractor}}
\label{app:sex}

A popular image segmentation tool in astronomy is \texttt{SExtractor}\footnote{\url{https://sextractor.readthedocs.io/en/latest/index.html}} \citep{1996A&AS..117..393B}. This software is largely used for the detection of astronomical sources, background reduction, and photometry of astronomical images, being especially suitable for processing large field-of-view images. However, 
%implementing 
running \texttt{SExtractor} can be computationally expensive, especially compared to sigma clipping. We decided to compare the performance of both methods considering only those events that have a spectral classification available on the Transient Name Server (TNS)\footnote{TNS is the International Astronomical Union's official mechanism for reporting new astronomical transients since 2016, \url{https://www.wis-tns.org/}.}. We find that \texttt{SExtractor} retrieves 149 hostless candidates while sigma clipping retrieves 181 hostless candidates. By visually inspecting each candidate in the search for the presence of a potential host, we find that the \texttt{SExtractor} method has a $\sim$ 15\% contamination while the sigma clipping method has a $\sim$ 22\% contamination. Thus, considering that running \texttt{SExtractor} involves writing and reading files on the disk, which is not ideal when working with large volumes of data; and that the performance of both methods is similar, we favor the simpler sigma clipping as an image segmentation method.

\clearpage
\section{Machine learning classified hostless candidates}
\label{app:ml_hc}
\begin{minipage}{0.5\textheight} 
\begin{ThreePartTable}
\begin{TableNotes}
      \item First column presents the IAU name of each object. The second column shows the corresponding ZTF internal name. The third and fourth column show right ascension and declination respectively. Column five shows the classification available on TNS.
\end{TableNotes}
\onecolumn
\begin{longtable}{llccl}
%\centering
\caption{Fragment of the hostless candidate list without a reported classification on TNS. The full list can be found as supplementary material.}
    \label{tab:TNSsample}
   \endfirsthead
    \hline\hline          % inserts double horizontal lines
 IAU Name & ZTF Name  & R.A.     &  Dec.       & Class    \\  
          &           & [J2000]  &  [J2000]    &          \\  
\hline  
AT~2016ayj  & \href{https://fink-portal.org/ZTF19adehksw}{ZTF19adehksw} & 03:06:45.60 &  46:09:12.93  & SN*\_cand.      \\
AT~2016ayl  & \href{https://fink-portal.org/ZTF18acwwwsg}{ZTF18acwwwsg} & 05:14:00.67 &  55:21:57.81  & SN*\_cand.      \\
AT~2016azn  & \href{https://fink-portal.org/ZTF22abxlizh}{ZTF22abxlizh} & 10:06:54.81 & -14:25:37.80  & AT~cand.	\\
AT~2017kn   & \href{https://fink-portal.org/ZTF18aakpggd}{ZTF18aakpggd} & 11:54:19.60 &  57:57:50.77  & QSO	        \\
AT~2018aod  & \href{https://fink-portal.org/ZTF23abofayp}{ZTF23abofayp} & 03:25:09.82 &  48:50:19.95  & SN~cand.	\\
AT~2018cou  & \href{https://fink-portal.org/ZTF18acxcpmo}{ZTF18acxcpmo} & 14:15:23.73 & -20:00:54.17  & SN	        \\
AT~2018ctv  & \href{https://fink-portal.org/ZTF18abtgnsi}{ZTF18abtgnsi} & 01:25:52.40 & -01:22:01.66  & SN	        \\
AT~2018cyo  & \href{https://fink-portal.org/ZTF19aavprpy}{ZTF19aavprpy} & 22:11:56.27 & -04:41:40.50  & SN	        \\
AT~2018fou  & \href{https://fink-portal.org/ZTF18abtefbi}{ZTF18abtefbi} & 23:05:32.51 &  00:49:02.50  & SN	        \\
AT~2018his  & \href{https://fink-portal.org/ZTF22abiflxl}{ZTF22abiflxl} & 17:49:31.59 &  17:15:37.23  & SN~cand.	\\
$\cdots$ & $\cdots$ & $\cdots$ & $\cdots$ & $\cdots$ \\
\hline
%\insertTableNotes
\end{longtable}
\twocolumn
\end{ThreePartTable}
\end{minipage}
%%%%%%%%%%%%%%%%%%%%%%%%%%%%%%%%%%%%%%%%%%%%%%%%%%

\end{document}